\documentclass[groupedaddress,
twocolumn,
showkeys, preprintnumbers,
amsmath,amssymb, aps, pre,
draft,
  floatfix,]{revtex4-1}

\usepackage{dynlearn}
\usepackage{subfig}
\usepackage{standalone}
\usepackage{amsthm}
\usepackage{mathrsfs}
\usepackage{graphicx}\usepackage{dcolumn}\usepackage{bm}
\theoremstyle{definition}

\newtheorem{theorem}{Theorem}

\renewcommand{\H}{\operatorname{H}} 

\newcommand{\cs}{\causalstate}

\newcommand{\One}{ {\mathbf{1} } }
\newcommand{\MSym}{\MeasSymbol}
\newcommand{\msym}{\meassymbol}

\newcommand{\MxSSet}{\AlternateStateSet}
\newcommand{\MxSMeasure}{\mu}
\newcommand{\MxSMeasureAlt}{\nu}
\newcommand{\MxSDyn}{\mathcal{W}}
\newcommand{\mxst}{\eta}
\newcommand{\mxstalt}{\zeta}
 \newcommand{\StartMS}{\bra{\delta_\pi}}

\newcommand{\simplex}{\Delta}
\newcommand{\ctvty}{ \tau }
\newcommand{\CoarseGraining}{\mathcal{C}}

\begin{document}

\title{Shannon Entropy Rate of Hidden Markov Processes}

\author{Alexandra M. Jurgens}
\email{amjurgens@ucdavis.edu}

\author{James P. Crutchfield}
\email{chaos@ucdavis.edu}

\affiliation{Complexity Sciences Center, Physics Department\\
University of California at Davis\\
Davis, California 95616}	

\date{\today}
\bibliographystyle{unsrt}

\begin{abstract}
Hidden Markov chains are widely applied statistical models of stochastic
processes, from fundamental physics and chemistry to finance, health, and
artificial intelligence. The hidden Markov processes they generate are
notoriously complicated, however, even if the chain is finite state: no finite
expression for their Shannon entropy rate exists, as the set of their
predictive features is generically infinite. As such, to date one cannot make
general statements about how random they are nor how structured. Here, we
address the first part of this challenge by showing how to efficiently and
accurately calculate their entropy rates. We also show how this method gives
the minimal set of infinite predictive features. A sequel addresses the
challenge's second part on structure.
\end{abstract}


\preprint{\arxiv{2008.XXXXX}}

\maketitle

\section{Introduction}
\label{sec:introduction}

Randomness is as necessary to physics as determinism. Indeed, since Henri
Poincar\'e's failed attempt to establish the orderliness of planetary motion,
it has been understood that both determinism and randomness are essential and
unavoidable in the study of physical systems \cite{Goro91a, Goro93a, Goro93b,
Crut86}. In the 1960s and 1970s, the rise of dynamical systems theory and the
exploration of statistical physics of critical phenomena offered up new
perspectives on this duality. The lesson was that intricate structures in a
system's state space amplify uncertainty, guiding it and eventually installing
it---paradoxically---in complex spatiotemporal patterns. Accepting this state
of affairs prompts basic, but as-yet unanswered questions. How is this
emergence monitored?  How do we measure a system's randomness or quantify its
patterns and their organization?

The tools needed to address these questions arose over recent decades during
the integration of Turing's computation theory \cite{Turi36, Shan56c, Mins67},
Shannon's information theory \cite{Shan48a}, and Kolmogorov's dynamical systems
theory \cite{Kolm56b, Kolm65, Kolm83, Kolm59, Sina59}. This established the
vital role that information plays in physical theories of complex systems. In
particular, the application of hidden Markov chains to model and analyze the
randomness and structure of physical systems has seen considerable success, not
only in complex systems \cite{Crut12a}, but also in coding theory
\cite{Marc11a}, stochastic processes \cite{Ephr02a}, stochastic thermodynamics
\cite{Bech15a}, speech recognition \cite{Rabi86a}, computational biology
\cite{Birney01, Eddy04}, epidemiology \cite{Breto2009}, and finance
\cite{Ryden98}, to offer a nonexhaustive list of examples.

A highly useful property of certain hidden Markov chains (HMCs) is
\emph{unifilarity} \cite{Ash65a}, a structural constraint on their state
transitions. Shannon showed that given a process generated by a finite-state
unifilar HMC, one may directly and accurately calculate a process' irreducible
randomness \cite{Shan48a}---now called the \emph{Shannon entropy rate}.
Furthermore, for such a process, there is a unique minimal finite-state
unifilar HMC that generates the process \cite{Crut88a}, known as the
\emph{\eM}. The \eM states---the process' \emph{causal states}---are the
minimal set of maximally predictive features. One consequence of the \eM's
uniqueness and minimality is that its mathematical description gives
a constructive definition of a process' structural complexity as the amount of
memory required to generate the process.

Loosening the unifilar constraint to consider a wider class of generated
processes, however, leads to major roadblocks. Predicting a process
generated by a \emph{finite-state} nonunifilar HMC requires an \emph{infinite
set} of causal states \cite{Crut92c}. That is, though ``finitely'' generated,
the process cannot be predicted by any finite unifilar HMC. Practically, this
precludes directly determining the process' entropy rate using Shannon's result
and, at best, obscures any insight into its internal structure.

That said, its causal states are (in general, see \cref{app:MinimalityofMSP})
equivalent to the uncountable set of \emph{mixed states}, or predictive
features, formally introduced by Blackwell over a half century ago
\cite{Blac57b}. To date, working with infinite mixed-states required
coarse-graining to produce a finite set of predictive features. Fortunately,
the tradeoffs between resource constraints and predictive power induced by such
coarse graining can be systematically laid out \cite{creutzig2009past, Stil07b,
Marz14f}.

The following introduces an alternative and more direct approach to working
with mixed states, though. It casts generating mixed states as a chaotic
dynamical system---specifically, a (place dependent) \emph{iterated function
system} (IFS). This obviates analyzing the underlying HMC via coarse graining.
Rather, the complex dynamics of the new system directly captures the
information-theoretic properties of the original process. Specifically, this
allows exactly calculating the entropy rate of the process generated by the
original nonunifilar finite-state HMC. Additionally, the IFS interpretation of
the nonunifilar HMC provides new insight into the structure and complexity of
infinite-state processes. This has direct application to the study of
randomness and structure in a wide range of physical systems.

In point of fact, the following and its sequel \cite{Jurg20c} were proceeded by
two companions that applied the theoretical results here to two, rather
different, physical domains. The first analyzed the origin of randomness and
structural complexity engendered by quantum measurement \cite{Vene19a}. The
second solved a longstanding problem on exactly determining the thermodynamic
functioning of Maxwellian demons, aka information engines \cite{Jurg20a}. That
is, the following and its sequel lay out the mathematical and algorithmic tools
required to successfully analyze these applied problems. We believe the new
approach is destined to find even wider applications.

Section \ref{sec:HiddenMarkovProcesses} recalls the necessary background in
stochastic processes, hidden Markov chains, and information theory. Section
\ref{sec:IteratedFunctionSystem} reviews the needed results on iterated
function systems; while Sec. \ref{sec:MixedStatePresentation} develops mixed
states and their dynamic---the mixed-state presentation. The main result
connecting these then follows in Sec. \ref{sec:MixedStateasanIFS}, showing that
the mixed-state presentation is an IFS and that it produces an ergodic process.
Section \ref{sec:entropygeneralHMC} recalls Blackwell's theory, updating it for
our present purpose of determining the entropy rate of any HMC. The
Supplementary Materials provide background on the asymptotic equipartition
property and minimality of the mixed states. They also constructively work
through the results for several example nonunifilar HMCs. They close with the
statistical error analysis underlying entropy-rate estimation.

\section{Hidden Markov Processes}
\label{sec:HiddenMarkovProcesses}

A \emph{stochastic process} $\Process$ is a probability measure over a
bi-infinite chain $\ldots \, \MSym_{t-2} \, \MSym_{t-1} \, \MSym_{t} \,
\MSym_{t+1} \, \MSym_{t+2} \ldots$ of random variables, each denoted by a
capital letter. A particular \emph{realization} $\ldots \, \msym_{t-2} \,
\msym_{t-1} \, \msym_{t} \, \msym_{t+1} \, \msym_{t+2} \ldots$ is denoted via
lowercase letters. We assume values $\msym_{t}$ belong to a discrete alphabet
$\MeasAlphabet$. We work with blocks $\MS{t}{t^\prime}$, where the first index
is inclusive and the second exclusive: $\MS{t}{t^\prime} = \MSym_{t} \ldots
\MSym_{t^\prime-1}$. $\Process$'s measure is defined via the collection of
distributions over blocks: $\{ \Pr(\MS{t}{t^\prime}): t < t^\prime, t,t^\prime
\in \mathbb{Z} \}$.

To simplify the development, we restrict to stationary, ergodic processes:
those for which $\Prob(\MS{t}{t+\ell}) = \Prob(\MS{0}{\ell})$ for all $t \in
\mathbb{Z}$, $\ell \in \mathbb{Z}^+$. In such cases, we only need to consider a
process's length-$\ell$ \emph{word distributions} $\Prob(\MS{0}{\ell})$.

A \emph{Markov process} is one for which $\Pr(\MSym_t|\MS{-\infty}{t}) =
\Pr(\MSym_t|\MSym_{t-1})$. A \emph{hidden Markov process} is the output of a
memoryless channel \cite{Cove06a} whose input is a Markov process
\cite{Ephr02a}. Working with processes directly is cumbersome, so we turn to
consider finitely-specified mechanistic models that generate them.

\begin{Def}
\label{Def:HMM}
A finite-state edge-labeled \emph{hidden MC} (HMC) consists of:
\begin{enumerate}
\setlength{\topsep}{0mm}
\setlength{\itemsep}{0mm}
\item a finite set of states $\CausalStateSet = \{\causalstate_1, ... ,
 \causalstate_N \}$,
\item a finite alphabet $\MeasAlphabet$ of $k$ symbols $x \in
 \MeasAlphabet$, and
\item a set of $N$ by $N$ symbol-labeled transition matrices $T^{(\msym)}$,
 $\msym \in \MeasAlphabet$: $T^{(\msym)}_{ij} =
 \Pr(\causalstate_j,\msym|\causalstate_i)$. The corresponding overall
 state-to-state transitions are described by the row-stochastic  matrix
 $T = \sum_{x \in \MeasAlphabet} T^{(x)}$.
\end{enumerate}
\end{Def}

\begin{figure}
 \centering
 \includegraphics{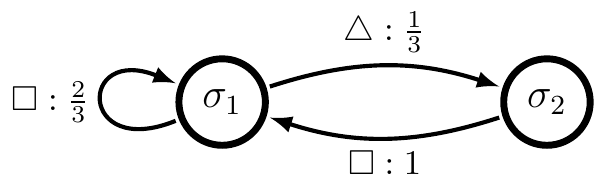}
 \caption[text]{A hidden Markov chain (HMC) with two states, $\{\sigma_1,
 \sigma_2 \}$ and two symbols $\{\square, \triangle\}$. This machine is
 unifilar.}
 \label{Fig:hiddenmarkovchain}
\end{figure}

Any given stochastic process can be generated by any number of HMCs. These are
called a process' \emph{presentations}.

We now introduce a structural property of HMCs that has important consequences
in characterizing process randomness and structure.

\begin{Def}
 A \emph{unifilar HMC} (uHMC) is an HMC such that for each state
 $\causalstate_i \in \CausalStateSet$ and each symbol $\msym \in
 \MeasAlphabet$ there is at most one outgoing edge from state
 $\causalstate_i$ labeled with symbol $\msym$.
\label{def:UHMC}
\end{Def}

Although there are many presentations for a process $\Process$, there is a
canonical presentation that is unique: a process' \emph{\eM}.

\begin{Def}
An \emph{\eM} is a uHMC with \emph{probabilistically distinct states}: For each
pair of distinct states $\causalstate_i, \causalstate_j \in \CausalStateSet$
there exists a finite word $w = \ms{0}{\ell-1}$ such that: 
\begin{align*}
\Prob(\MS{0}{\ell} = w|\CausalState_0 = \causalstate_k)
  \not= \Prob(\MS{0}{\ell} = w|\CausalState_0 = \causalstate_j)~.
\end{align*}
\label{def:eM}
\end{Def}

A process' \eM\ is its optimal, minimal presentation, in the sense that the
set of predictive states |$\CausalStateSet|$ is minimal compared to all its
other unifilar presentations \cite{Shal98a}.

\subsection{Entropy Rate of HMCs}
\label{subsec:EntropyRateHMCs}

A process' intrinsic randomness is the information in the present measurement,
discounted by having observed the information in an infinitely long
history. It is measured by Shannon's source entropy rate \cite{Shan48a}.

\begin{Def}
\label{Def:EntropyRate}
A process' \emph{entropy rate} $\hmu$ is the asymptotic average
entropy per symbol \cite{Crut01a}: 
\begin{align}
\hmu = \lim_{\ell \to \infty} \H[\MS{0}{\ell}] / \ell
  ~,
\label{eq:EntropyDensity}
\end{align}
where $H[\MS{0}{\ell}]$ is the Shannon entropy of block $\MS{0}{\ell}$:
\begin{align}
H[\MS{0}{\ell}] = - \sum_{\ms{0}{\ell} \in \MeasAlphabet^\ell}
  \Pr(\ms{0}{\ell}) \log_2 \Pr(\ms{0}{\ell})
  ~.
\label{eq:Shannonentropy}
\end{align}
\end{Def}

Given a finite-state unifilar presentation $M_u$ of a process $\Process$, we may
directly calculate the entropy rate from the transition matrices of the uHMC
\cite{Shan48a}:
\begin{align}
\hmu(\Process) & = \hmu(M_u) \nonumber \\
  & = - \sum_{\cs \in \CausalStateSet} \Pr(\cs)
  \sum_{\msym \in \MeasAlphabet} T_{\cs \cs^\prime}^{(\msym)}
  \log_2 T_{\cs \cs^\prime}^{(\msym)}
  ~.
\label{eq:ShannonEntropyRate}
\end{align}
Blackwell showed, though, that in general for processes generated by HMCs there
is no closed-form expression for the entropy rate \cite{Blac57b}. For a process
generated by an nonunifilar HMC $M$, applying \cref{eq:ShannonEntropyRate} to
$M$ typically overestimates the true entropy rate of the process
$\hmu(\Process)$:
\begin{align*}
\hmu(M) \geq \hmu(\Process)
  ~.
\end{align*}
Overcoming this limitation is one of our central results.
We now embark on introducing the necessary tools for this.

\section{Iterated Function Systems}
\label{sec:IteratedFunctionSystem}

To get there, we must take a short detour to review iterated function systems
(IFSs) \cite{Barn88a}, as they play a critical role in analyzing HMCs.
Speaking simply, we show that HMCs are dynamical systems---namely, IFSs.

Let $(\simplex^N, d)$ be a compact metric space with $d(\cdot,\cdot)$ a
distance. This notation anticipates our later application, in which
$\simplex^N$ is $N$-simplex of discrete-event probability distributions (see
\cref{sec:MixedStates}). However, the results here are general.

Let $f^{(\msym)}: \simplex^N \to\simplex^N$ for $x = 1, \dots, k$ be a set of
Lipschitz functions with:
\begin{align*}
d\left(f^{(\msym)} (\mxst), f^{(\msym)} (\mxstalt) \right)
  \leq \ctvty^{(\msym)} d(\mxst, \mxstalt)
  ~,
\end{align*}
for all $\mxst, \mxstalt \in \simplex^N$ and where $\ctvty^{(\msym)}$ is a
constant. This notation is chosen to draw an explicit parallel to the
stochastic processes discussed in \cref{sec:HiddenMarkovProcesses} and to avoid
confusion with the lowercase Latin characters used for realizations of
stochastic processes. In particular, note that the superscript ${(\msym)}$ here
and elsewhere parallels that of the HMC symbol-labeled transition matrices
$T^{(\msym)}$. The reasons for this will soon become clear.

The Lipschitz constant $\ctvty^{(\msym)}$ is the \emph{contractivity} of map
$f^{(\msym)}$. Let $p^{(\msym)} : M \to [0,1]$ be continuous, with
$p^{(\msym)}(\mxst) \geq 0$ and $\sum_{\msym=1}^k p^{(\msym)}(\mxst) = 1$ for
all $\mxst$ in $M$. The triplet $\{\simplex^N,\{p^{(x)}\},\{f^{(x)}\}: x \in
\MeasAlphabet\}$ defines a \emph{place-dependent} IFS.

A place-dependent IFS generates a stochastic process over $\eta \in \simplex^N$
as follows. Given an initial position $\mxst_0 \in \simplex^N$, the probability
distribution $\{ p^{(\msym)}(\mxst_0) : \msym = 1, \dots, k \}$ is sampled.
According to the sample $\msym$, apply $f^{(\msym)}$ to map $\mxst_0$ to the
next position $\mxst_1 = f^{(\msym)}(\mxst_0)$. Resample $x$ from the
distribution and continue, generating $\mxst_0, \mxst_1, \mxst_2, \ldots$.

If each map $f^{(\msym)}$ is a contraction---i.e., $\ctvty^{(\msym)} < 1$
for all $\mxst, \mxstalt \in \simplex^N$---it is well known that there
exists a unique nonempty compact set $\Lambda \subset \simplex^N$ that is
invariant under the IFS's action:
\begin{align*}
\Lambda = \bigcap_{\msym=1}^k f^{(\msym)}(\Lambda)
  ~.
\end{align*}
$\Lambda$ is the IFS's \emph{attractor}.

Consider the operator $V : M(\simplex^N) \to M(\simplex^N)$ on the space of
Borel measures on the $N$-simplex: 
\begin{align}
V \MxSMeasure (B)
  = \sum^k_{\msym =1}
  \int_{\left(f^{(\msym)}\right)^{-1} (B) }
  p^{(\msym)}(\mxst) d \MxSMeasure (\mxst)
  ~.
\label{eq:IFSOperator}
\end{align}
A Borel probability measure $\MxSMeasure$ is said to be \emph{invariant} or
\emph{stationary} if $V\MxSMeasure = \MxSMeasure$. It is \emph{attractive}
if for any probability measure $\MxSMeasureAlt$ in $M(\simplex^N)$:
\begin{align*}
\int g d(V^n \MxSMeasureAlt) \to \int g \MxSMeasure
  ~,
\end{align*}
for all $g$ in the space of bounded continuous functions on $\simplex^N$. 

Let's recall here a key result concerning the existence of attractive,
invariant measures for place-dependent IFSs.

\begin{theorem}
\cite[Thm. 2.1]{Barns88} Suppose there exists $r<1$ and $q>0$ such that:
\begin{align*}
  \sum_{x\in\MeasAlphabet} p^{(\msym)} (\mxst) d^q
  \left(f^{(\msym)} (\mxst), f^{(\msym)} (\mxstalt) \right)
  \leq r^q d^q \left(\mxst, \mxstalt \right)
  ~,
\end{align*}
for all $\mxst, \mxstalt \in \simplex^N$. Assume that the modulus of uniform
continuity of each $p^{(x)}$ satisfies Dini's condition and that there exists a
$\delta > 0$ such that:
\begin{align}
  \sum_{x: d(f^{(\msym)}(\mxst), f^{(\msym)}(\mxstalt)) \leq r d(\mxst, \mxstalt)}
  p^{(\msym)}(\mxst) p^{(\msym)}(\mxstalt) \leq \delta^2 
  ~,
\end{align}
for all $\mxst, \mxstalt \in \simplex^N$. Then there is an attractive, unique,
invariant probability measure for the Markov process generated by the
place-dependent IFS.
\label{theorem:PDIFS_Measure}
\end{theorem}

In addition, under these same conditions Ref. \cite{Elton1987} established an
ergodic theorem for IFS \emph{orbits}. That is, for any $\mxst \in \simplex^N$
and $g: \simplex^N \to \simplex^N$:
\begin{align}
\frac{1}{n+1} \sum_{k=0}^n
  g(w_{\msym_k} \circ \dots \circ w_{\msym_1} \mxst ) \to \int g d \MxSMeasure 
  ~.
\label{eq:IFSergodic}
\end{align}

\section{Mixed-State Presentation}
\label{sec:MixedStatePresentation}

We now return to stochastic processes and their HMC presentations. When
calculating entropy rates from various presentations, we noted that HMC
presentations led to difficulties: (i) the internal Markov-chain entropy-rate
overestimates the process' entropy rate and (ii) there is no closed-form
entropy-rate expression. To develop the tools needed to resolve these problems,
we introduce HMC \emph{mixed states} and their dynamic.

Assume that an observer has a finite HMC presentation $M$ for a process
$\Process$. Since the process is hidden, the observer does not directly measure
$M$'s internal states. Absent output data, the best guess for $M$'s hidden
states is that they occur according to the state stationary distribution $\pi$.
The observer can improve on this guess by monitoring the output data $\msym_{0}
\, \msym_{1} \, \msym_{2} \ldots$ that $M$ generates. Given knowledge of $M$,
determining the internal state from observed data is the problem of
\emph{observer-process synchronization}.

\subsection{Mixed States}
\label{sec:MixedStates}

For a length-$\ell$ word $w$ generated by $M$ let $\mxst(w) =
\Prob(\CausalStateSet|w)$ be the observer's \emph{belief distribution} as
to the process' current state after observing $w$: 
\begin{align}
 \mxst(w) & \equiv \Prob(\CausalState_\ell | \MS{0}{\ell}=w, \CausalState_0 \sim \pi) ~.
\label{eq:MixedState}
\end{align}
When observing a $N$-state machine, the vector $\bra{\mxst(w)}$ lives in the
\emph{(N-1)-simplex} $\simplex^{N-1}$, the set such that:
\begin{align*}
 \{ \eta \in \mathbb{R}^{N} : \langle\eta\ket{\One} = 1, \langle\eta\ket{\delta_i} \geq 0,
  i=1, \dots , N \}
  ~,
\end{align*}
where $\bra{\delta_i} = \begin{pmatrix} 0 & 0 & \dots & 1 & \dots & 0
\end{pmatrix}$. The $0$-simplex $\simplex^0$ is the single point $\ket{\eta} =
(1)$, the $1$-simplex $\simplex^1$ is the line segment $[0,1]$ from $\ket{\eta}
= (0,1)$ to $\ket{\eta} = (1,0)$, and so on.  

The set of belief distributions $\mxst(w)$ that an HMC can visit defines
its set of \emph{mixed states}:
\begin{align*}
 \MxSSet = \{ \mxst(w): w \in \MeasAlphabet^+, \Pr(w) > 0 \} ~.
\end{align*}
Generically, the mixed-state set $\MxSSet$ for an $N$-state HMC is infinite,
even for finite $N$ \cite{Blac57b}. 

Note that when a mixed state appears in probability expressions, the notation
refers to the random variable $\mxst$, not the row vector $\ket{\mxst}$, and we
drop the bra-ket notation. Bra-ket notation is used in vector-matrix
expressions.

\subsection{Mixed-State Dynamic}
\label{sec:MixedStateMethod}

The probability of transitioning from $\bra{\mxst(w)}$ to $\bra{\mxst(w\msym)}$
on observing symbol $\msym$ follows from \cref{eq:MixedState} immediately; we
have:
\begin{align*}
\Pr(\mxst(w\msym) | \mxst(w)) = \Pr(\msym|\CausalState_\ell \sim \mxst(w))
 ~.
\end{align*}
This defines the mixed-state transition dynamic $\MxSDyn$. Together the mixed
states and their dynamic define an HMC that is unifilar by construction. This is
a process' \emph{mixed-state presentation} (MSP) $\MSP(\Process) = \{\MxSSet, \MxSDyn \}$.

We defined a process' $\MSP$ abstractly. The $\MSP$ typically has an uncountably
infinite set of mixed states, making it challenging to work with in the form
laid out in \cref{sec:MixedStates}. Usefully, however, given any HMC $M$ that
generates the process, we may explicitly write down the dynamic $\MxSDyn$.
Assume we have an $N+1$-state HMC presentation $M$ with $k$ symbols $\msym \in
\MeasAlphabet$. The initial condition is the invariant probability $\pi$ over
the states of $M$, so that $\bra{\mxst_0} = \StartMS$. In the context of the
mixed-state dynamic, mixed-state subscripts denote time.
 
\begin{figure}
\centering
\includegraphics[width=.8\columnwidth]{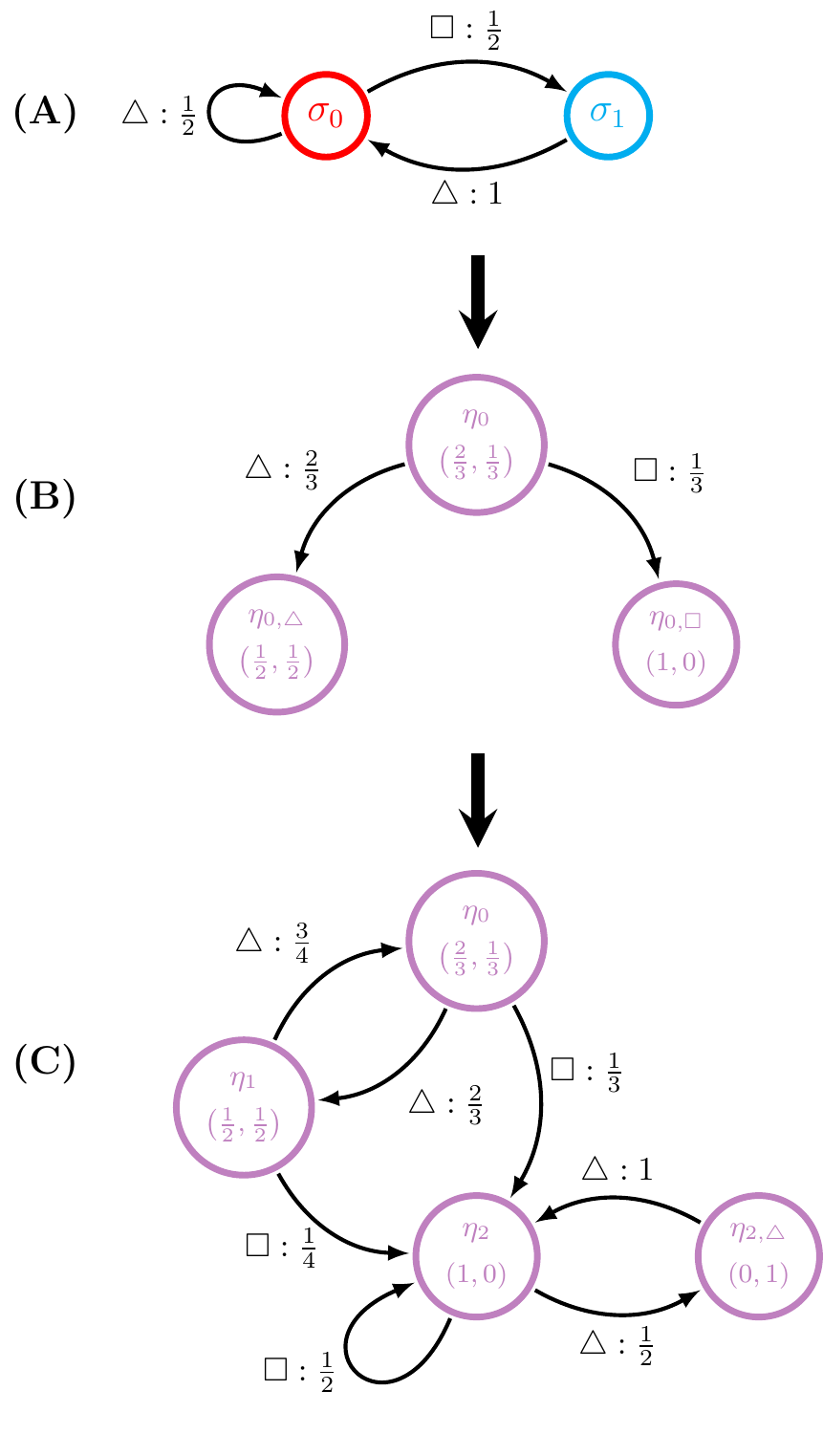}
\caption[text]{Determining the mixed-state presentation (MSP) of the $2$-state
	unifilar HMC shown in (A): The invariant state distribution $\pi = (2/3,
	1/3)$. It becomes the first mixed state $\eta_0$ used in (B) to calculate
	the next set of mixed states. (C) The full set of mixed states seen from
	all allowed words. In this case, we recover the unifilar HMC shown in (A)
	as the MSP's recurrent states.
	}
\label{Fig:mixedstatemethod}
\end{figure}
 
The probability of generating symbol $\msym$ when in mixed state $\mxst$ is:
\begin{align}
  \Pr(\msym|\mxst) = \bra{\mxst} T^{(\msym)} \ket{\One}
  ~,
\label{eq:SymbolFromMixedState}
\end{align}
where $T^{(\msym)}$ is the symbol-labeled transition matrix associated with the
symbol $\msym$.

From $\mxst_0$, we calculate the probability of seeing each $\msym \in
\MeasAlphabet$. Upon seeing symbol $\msym$, the current mixed state
$\bra{\mxst_t}$ is updated according to:
\begin{align}
  \bra{\mxst_{t+1, \msym}}
  = \frac{\bra{\mxst_t} T^{(\msym)}} {\bra{\mxst_t} T^{(\msym)} \ket{\One}}
  ~.
\label{eq:MxStUpdate}
\end{align}

\begin{figure*}
\centering
\includegraphics[width=.85\textwidth]{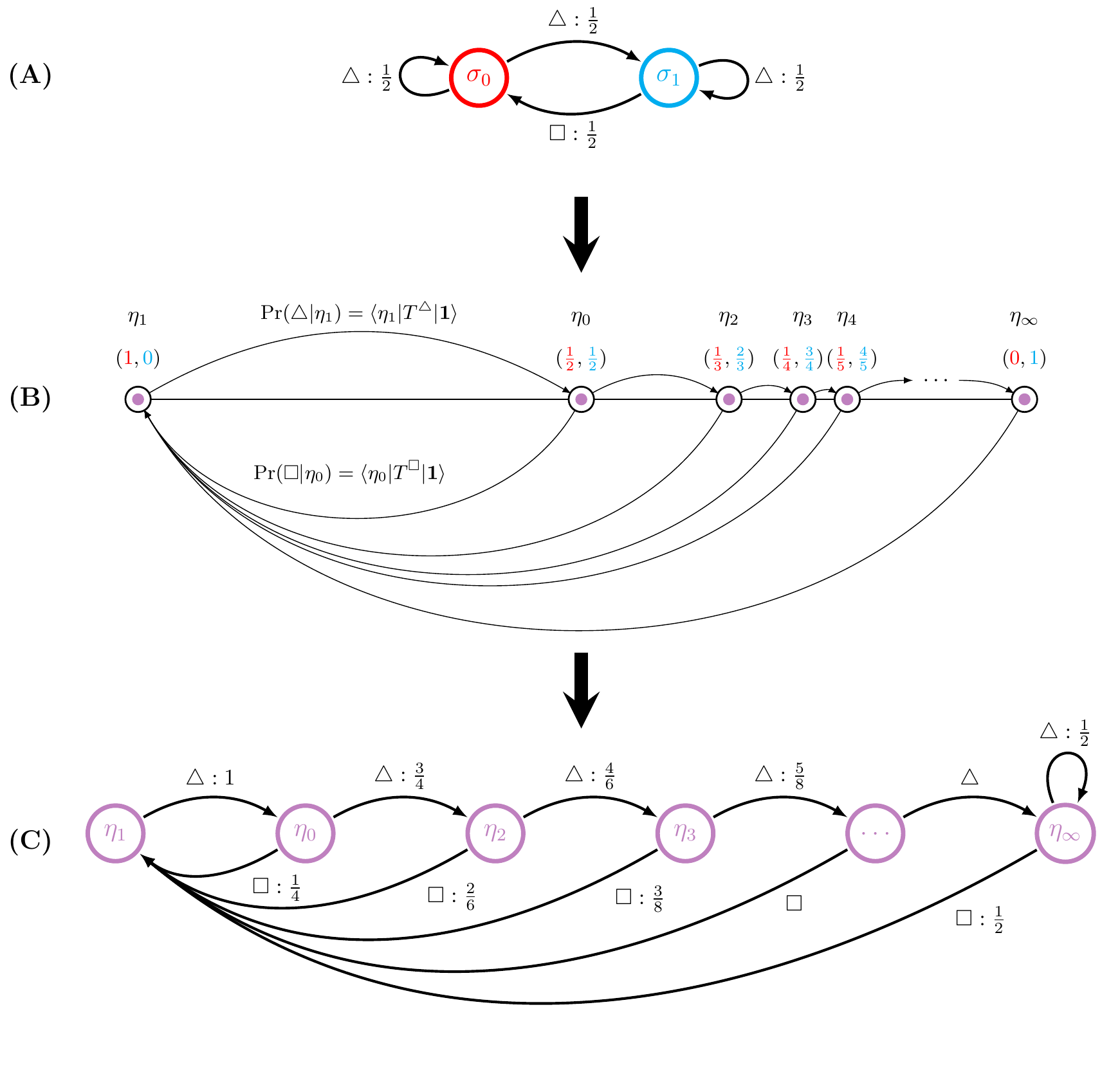}
\caption[text]{Determining the mixed-state presentation of the $2$-state
	\emph{nonunifilar} HMC shown in (A). The invariant distribution $\pi =
	(1/2, 1/2)$. It is the first mixed state $\eta_0$ used in
	(B) to calculate the next set of mixed states. (B) plots the mixed states
	along the $1$-simplex $\simplex^1 = [0,1]$. In (C), we translated the
	points on the simplex to the states of an infinite-state, unifilar HMC.
	}
\label{Fig:simplextomachine}
\end{figure*}
 
Thus, given an HMC presentation we can restate \cref{eq:MixedState} as:
\begin{align*}
\bra{\mxst(w)} & = \frac{\bra{\mxst_0} T^{(w)}}{\bra{\mxst_0} T^{(w)}
\ket{\One}} \\
  & = \frac{\bra{\pi} T^{(w)}}{\bra{\pi} T^{(w)} \ket{\One}}
  ~.
\end{align*}
Equation (\ref{eq:MxStUpdate}) tells us that, by construction, the MSP is
unifilar, since each possible output symbol uniquely determines the next (mixed)
state. Taken together, \cref{eq:SymbolFromMixedState,eq:MxStUpdate} define the
mixed-state transition dynamic $\MxSDyn$ as:
\begin{align*}
\Pr(\mxst_{t+1},\msym|\mxst_t) & = \Pr(\msym|\mxst_t) \\
                               & = \bra{\mxst_t} T^{(\msym)} \ket{\One} 
  ~,
\end{align*}
for all $\mxst \in \MxSSet$, $\msym \in \MeasAlphabet$.

To find the MSP $\MSP = \{\MxSSet, \MxSDyn \}$ for a given HMC $M$ we apply
the \emph{mixed-state construction method}:
\begin{enumerate}
\setlength{\topsep}{-3mm}
\setlength{\itemsep}{-1mm}
\item Set $\MSP = \{ \MxSSet = \emptyset, \MxSDyn = \emptyset \}$.
\item Calculate $M$'s invariant state distribution: $\pi = \pi T$.
\item Take $\mxst_0$ to be $\StartMS$ and add it to $\MxSSet$.
\item For each current mixed state $\mxst_t \in \MxSSet$, use
	\cref{eq:SymbolFromMixedState} to calculate $\Pr(\msym | \mxst_t )$ for
	each $\msym \in \MeasAlphabet$. 
\item For $\mxst_t \in \MxSSet$, use \cref{eq:MxStUpdate} to find the updated
	mixed state $\mxst_{t+1, \msym}$ for each $\msym \in \MeasAlphabet$. 
\item Add $\mxst_{t}$'s transitions to $\MxSDyn$ and each $\mxst_{t+1, x}$
	to $\MxSSet$, merging duplicate states. 
\item For each new $\mxst_{t+1}$, repeat steps 4-6 until no new mixed states
	are produced.
 \end{enumerate}
With the MSP $\MSP(M)$ in hand, the next issue is determining it's
(equivalent) \eM. There are several cases.

Beginning with a finite, unifilar HMC $M$ generating a process $\Process$, the
MSP $\MSP(M)$ is a finite, optimally-predictive rival presentation to
$\Process$'s \eM, as seen in \cref{Fig:mixedstatemethod}. In this case, the
starting HMC depicted in \cref{Fig:mixedstatemethod} (A) is an \eM, and
reducing the MSP in \cref{Fig:mixedstatemethod} (C) by trimming the transient
states returns the process' recurrent-state \eM. When starting with the \eM,
trimming the resultant $\MSP(\eM)$ in this way always returns the \eM.

In general, if $\MSP(M)$ is finite, we find the \eM by minimizing $\MSP(M)$ via
merging duplicate states: repeat mixed-state construction on $\MSP(M)$ and trim
transient states once more. Minimizing countably-infinite and
uncountably-infinite $\MSP(M)$ is discussed further in
\cref{app:MinimalityofMSP}.

The MSPs of unifilar presentations are interesting and contain additional
information beyond the unifilar presentations. For example, containing
transient causal states, they are employed in calculating many complexity
measures that track convergence statistics \cite{Crut13a}.

However, here we focus on the mixed-state presentations of nonunifilar HMCs,
which typically have an infinite mixed-state set $\MxSSet$.
\Cref{Fig:simplextomachine} illustrates applying mixed-state construction to a
finite, nonunifilar HMC. This produces an infinite sequence of mixed states on
$\simplex^1 =[0,1]$, as plotted in \cref{Fig:simplextomachine}(B). In this
particular example, the MSP is highly structured and $\MxSSet$ is countably
infinite, allowing us to better understand the underlying process $\Process$;
compared, say, to the $2$-state nonunifilar HMC in
\cref{Fig:simplextomachine}(A). MSPs of nonunifilar HMCs typically have an
uncountably-infinite mixed-state set $\MxSSet$.

\section{MSP as an IFS}
\label{sec:MixedStateasanIFS}

With this setup, our intentions in reviewing iterated function systems (IFSs)
become explicit. The mixed-state presentation (MSP) exactly defines a
place-dependent IFS, where the mapping functions are the set of symbol-labeled
mixed-state update functions as given in \cref{eq:MxStUpdate} and the set of
place-dependent probability functions are given by
\cref{eq:SymbolFromMixedState}. We then have a mapping function and associated
probability function for each symbol $\msym \in \MeasAlphabet$ that can be
derived from the symbol-labeled transition matrix $T^{(\msym)}$.

If these probability and mapping functions meet the conditions of
\cref{theorem:PDIFS_Measure}, we identify the attractor $\Lambda$ as the set
of mixed states $\MxSSet$ and the invariant measure $\MxSMeasure$ as the
invariant distribution $\pi$ of the potentially infinite-state $\MSP$. This
is the original HMC's Blackwell measure. Since all Lipschitz continuous
functions are Dini continuous, the probability functions meet the conditions by
inspection. We now establish that the maps are contractions, by appealing to
Birkhoff's 1957 proof that a positive linear map preserving a convex cone is a
contraction under the Hilbert projection metric \cite{Birkhoff57}.

Given an integer $N \geq 2$, let $C^N$ be the nonnegative cone in
$\mathbb{R}^N$, so that $C^N$ consists of all vectors $z = (z_1, z_2, \dots,
z_N)$ satisfying $z \neq 0$ and $z_i \geq 0$ for all $i$. The \emph{projective
distance} $d : C^N \times C^N \to [0, \infty)$ is defined:
\begin{align}
& d(z, y) := \nonumber \\
  & \quad\max \left\{ \left| \log \left( \frac{z_r}{z_s} \frac{y_s}{y_r} \right) \right| : r, s = 1, \dots, N ; r \neq s  \right\}
\label{eq:HilbertMetric}
\end{align}
for $z, y \in C^N$, where $d(z, z) = 0$. If one of the points is on the cone
boundary, the distance is taken to be $+\infty$. Note that the projective
distance, by construction, defines $d(\alpha z, \beta y) = d(z, y)$, where
$\alpha, \beta \in \mathbb{R}^+$. In other words, for two mixed states $\mxst,
\mxstalt \in \simplex^N$, $d \left( f^{(\msym)} ( \mxst) , f^{(\msym)}(
\mxstalt ) \right) = d ( \mxst T^{(\msym)}, \mxstalt T^{(\msym)} ) $.

If $T^{(\msym)}$ is an $N \times N$ positive matrix, we have $d( z
T^{(\msym)}, y T^{(\msym)}) < d_N(z, y)$ for every $z, y \in C^N$ such that
$d (y, z) > 0$. We define the \emph{projective contractivity} $\ctvty^{(\msym)}$
associated with $T^{(\msym)}$ as:
\begin{align*}
 \tau_\msym := \sup_{ \{ z, y \in C_N : d(z, y) > 0 \}  } \frac{d ( z T^{(\msym)}, y T^{(\msym)} )}{d(z,y)}
  ~,
\end{align*}
so that $\ctvty^{(\msym)}$ satisfies $\ctvty^{(\msym)} \leq 1$. As the
theorem below indicates, this inequality is strict.

\begin{theorem}
\label{theorem:contractioncoefficient}
(\cite[Thm. 1]{Cavazos03}.) Let the integers $m, n \geq 2$ be arbitrary. For
each matrix $T^{(\msym)} = \left[t^{(\msym)}_{ij}\right]$ of order $m \times n$
with positive components, $\ctvty^{(\msym)}$ is given by the following Birkhoff
formula:
\begin{align*}
     \ctvty^{(\msym)} = \frac{1 -\left( \phi^{(\msym)} \right)^{1/2}} {1 + \left( \phi^{(\msym)}\right)^{1/2}} ~,
\end{align*}
where:
\begin{align*}
     \phi(H) := \min_{r, s, j, k} \frac{t^{(\msym)}_{rj} t^{(\msym)}_{sk}}{t^{(\msym)}_{sj}t^{(\msym)}_{rk}}
  ~.
\end{align*}
\end{theorem}
By inspection we see that $\phi(H) > 0$ and $\ctvty < 1$. As Ref.
\cite{Kohlberg82} notes, not only does the projective metric turn all positive
linear transformations into contraction mappings, it is the only metric that
does so.

Positivity of the transition matrix guarantees that any boundary points are
mapped inside $\simplex^N$. This is not generally true for our transition
matrices---they are restricted merely to be nonnegative. However, the above
result extends to any nonnegative matrix $T^{(\msym)}$ for which there exists
an $N \in \mathbb{N}^+$ such that $\left( T^{(\msym)} \right)^N$ is a
positive matrix. Then there will be a $\ctvty^{(\msym)} < 1$ such that $d
\left( \mxst \left( T^{(\msym)} \right)^N, \mxstalt \left(T^{(\msym)} \right)^N
\right) < d (\mxst, \mxstalt)$. This is equivalent to a requirement that
$T^{(\msym)}$ be aperiodic and irreducible.

\begin{figure}[ht]
\centering
\includegraphics[width=.45\textwidth]{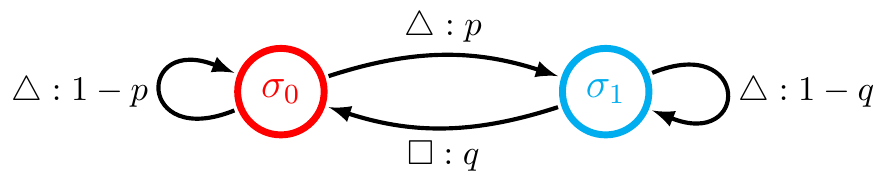}
\caption{Simple Nonunifilar Source (SNS): The symbol-labeled transition matrices
	given in \cref{eq:SNS_matrices} are both reducible, but the place-dependent
	IFS still has an attractor with an invariant probability distribution. By
	setting $p = q = 1/2$, we return the nonunifilar HMC from
	\cref{Fig:simplextomachine}.
  }
\label{fig:SNS}
\end{figure}

Still, we are not guaranteed irreducibility and aperiodicity for our
symbol-labeled transition matrices. Indeed, the \emph{Simple Nonunifilar
Source}, depicted in \cref{fig:SNS}, has the symbol-labeled transition
matrices:
\begin{align}
 T^{(\triangle)} = \begin{pmatrix}
     1-p & p  \\
     0 & 1-q
 \end{pmatrix}
 ~\text{and}~
 T^{(\square)} = \begin{pmatrix}
     0 & 0  \\
     q & 0
 \end{pmatrix}
~.
\label{eq:SNS_matrices}
\end{align}
Both $T^{(\triangle)}$ and $T^{(\square)}$ are reducible. A quick check is to
examine \cref{fig:SNS} and ask if there is a length-$n$ sequence consisting of
only a single symbol that reaches every state from every other state.
Nonetheless, the HMC has a countable set of mixed states $\MxSSet$ and an
invariant measure $\MxSMeasure$.

We can determine this from the mapping functions:
\begin{align}
f^{(\triangle)}(\mxst) & = \left[ \frac{ \langle \mxst \ket{\delta_1} 
 (1-p)}{1 - (1- \langle \mxst \ket{\delta_1} )q} , \right. \nonumber \\ 
 & \qquad \left. \frac{\langle \mxst \ket{\delta_1} p + 
 (1-\langle \mxst \ket{\delta_1})(1-q)}{1 + 
 (1-\langle \mxst \ket{\delta_1})q}
 \right] ~\text{and}~ \\
 f^{(\square)}(\mxst) & = \left[ 1, 0 \right]
 ~.
\label{eq:SNS_functions}
\end{align}
From any initial state $\mxst_0$, other than $\mxst_0 = \sigma_0 = \left[1,0
\right]$, the probability of seeing a $\square$ is positive. Once a $\square$ is
emitted, the mixed state is guaranteed to be $\mxst = \sigma_0 = \left[ 1, 0
\right]$. In this case, when the mapping function is constant and the
contractivity is $-\infty$, we call the symbol a \emph{synchronizing} symbol.
From $\sigma_0$, the set of mixed states is generated by repeated emissions of
$\triangle$s, so that $\MxSSet = \left\{\left ( f^{(\triangle) } \right)^n
(\sigma_0 ) : n = 0, \dots, \infty \right\}$. This is visually depicted in
\cref{Fig:simplextomachine} for the specific case of $p = q = 1/2$. For all
$p$ and $q$, the measure can be determined analytically; see Ref.
\cite{Marz14a}. Note that this is due to the HMC's highly structured topology.
In general, the set of mixed states is uncountable---either a fractal or
continuous set---and the measure cannot be analytically expressed.

Assuming the HMC generates an ergodic process ensures that the total transition
matrix $T = \sum_\msym T^{(\msym)}$ is nonnegative, irreducible, and aperiodic.
Define for any word $w = x_1 \ldots x_\ell \in \MeasAlphabet^+$ the associated
mapping function $T^{(w)} = T^{(x_1)} \circ \cdots \circ T^{(x_\ell)}$.
Consider word $w$ in a process' typical set of realizations (see
\cref{app:AEP}), which set approaches measure one as $\vert w \vert \to
\infty$. Due to ergodicity, it must be the case that $f^{(w)}$ is either (i) a
constant mapping---and, therefore, infinitely contracting---or (ii) $T^{(w)}$
is irreducible.

As an example of the former case, we see that any composition of the SNS
functions \cref{eq:SNS_functions} is always a constant function, so long as
there is at least one $\square$ in the word, the probability of which
approaches one as the word grows in length.

As an example of the later case, imagine adding to the SNS in \cref{fig:SNS} a transition on $\square$ from $\sigma_0$ to $\sigma_1$. Then, both symbol-labeled transition matrices are still reducible, but the composite transition matrices for any word including both symbols is now irreducible. Therefore, the map is contracting. While this is not the case for words composed of all $\square$s and all $\triangle$s, these sequences are measure zero as $N \to \infty$. \Cref{app:AEP} discusses this further.

\section{Entropy of General HMCs}
\label{sec:entropygeneralHMC}

Blackwell analyzed the entropy of \emph{functions of finite-state Markov chains}
\cite{Blac57b}. With a shift in notation, functions of Markov chains can be
identified as general hidden Markov chains. This is to say, both presentation
classes generate the same class of stochastic processes. As we have discussed,
the entropy rate problem for unifilar hidden Markov chains is solved, with
Shannon's entropy rate expression, \cref{eq:ShannonEntropyRate}. However,
according to Blackwell, there is no analogous closed-form expression for the
entropy rate of a nonunifilar HMC.

\subsection{Blackwell Entropy Rate}
\label{sec:blackwellentropy}

That said, Blackwell gave an expression for the entropy rate of general HMCs, by
introducing mixed states over stationary, ergodic, finite-state chains.
(Although he does not refer to them as such.) His main result, retaining his
notation, is transcribed here and adapted by us to constructively solve the
HMC entropy-rate problem. 

\begin{theorem}
(\cite[Thm. 1]{Blac57b}.) Let $\{ x_n, - \infty < n < \infty \}$ be a
stationary ergodic Markov process with states $i = 1, \dots, I$ and transition
matrix $M = \left\Vert m(i,j) \right\Vert$. Let $\Phi$ be a function defined on
$1, \dots, I$ with values $a = 1, \dots, A$ and let $y_n = \Phi(x_n)$. The
entropy of the $\{y_n\}$ process is given by:
\begin{align}
    H = - \int \sum_{a} r_a(w) \log r_a(w) d Q(w)
  ~,
\label{eq:Blackwellentropy}
\end{align}
where $Q$ is a probability distribution on the Borel sets of the set $W$ of
vectors $w = (w_1, \dots, w_I)$ with $w_i \geq 0$, $\sum_i w_i = 1$, and $r_a(w)
= \sum_{i=1}^I \sum_{j \owns \Phi(j) = a} w_i m(i,j)$. The distribution $Q$ is
concentrated on the sets $W_1, \dots, W_A$, where $W_a$ consists of all $w \in
W$ with $w_i = 0$ for $\Phi(i) \neq a$ and satisfies:
\begin{align}
    Q(E) = \sum_a \int_{f_a^{-1} E} r_a(w) d Q(w)
  ~,
\label{eq:Blackwellmeasure}
\end{align}
where $f_a$ maps $W$ into $W_a$, with the $j$th coordinate of $f_a(w)$ given by
$\sum_i w_i m(i,j) / r_a(w)$ for $\Phi(j) = a$.
\label{theorem:Blackwellentropy}
\end{theorem}

We can identify the $w$ vectors in \cref{theorem:Blackwellentropy} as exactly
the mixed states of \cref{sec:MixedStatePresentation}. Furthermore, it is clear
by inspection that $r_a(w)$ and $f_a(w)$ are the probability and mapping
functions of \cref{eq:SymbolFromMixedState,eq:MxStUpdate}, respectively, with
$a$ playing the role of our observed symbol $\msym$.

Therefore, Blackwell's expression \cref{eq:Blackwellentropy} for the HMC
entropy rate, in effect, replaces the average over a finite set
$\CausalStateSet$ of unifilar states in Shannon's entropy rate formula
\cref{eq:ShannonEntropyRate} with (i) the mixed states $\MxSSet$ and (ii) an
integral over the Blackwell measure $\MxSMeasure$. In our notation, we write
Blackwell's entropy formula as:
\begin{align}
 \hmu^B = - \int_{\MxSSet} d \mu(\mxst) \sum_{\msym \in \MeasAlphabet} p^{(\msym)}(\mxst)  \log_2  p^{(\msym)}(\mxst)
  ~.
\label{eq:Blackwellentropyintegral}
\end{align}

Thus, as with Shannon's original expression, this too uses unifilar
states---now, though, states from the mixed-state presentation $\MSP$. This, in
turn, maintains the finite-to-one internal (mixed-) state sequence to
observed-sequence mapping. Therefore, one can identify the mixed-state entropy
rate itself as the process' entropy rate.

\subsection{Calculating the Blackwell HMC Entropy}
\label{subsec:BlackwellCalculation}

Appealing to Ref. \cite{Elton1987}, we have that contractivity of our
substochastic transition matrix mappings guarantees ergodicity over the words
generated by the mixed-state presentation. And so, we can replace
\cref{eq:Blackwellentropyintegral}'s integral over $\MxSSet$ with a time
average over a mixed-state trajectory $\mxst_0, \mxst_1, \ldots$ determined by
a long allowed word, using Eqs. (\ref{eq:SymbolFromMixedState}) and
(\ref{eq:MxStUpdate}). This gives a new limit expression for the HMC entropy rate:
\begin{align}
 \widehat{\hmu}^B = - \lim_{\ell \to \infty} \frac{1}{\ell} \sum_{\msym \in \MeasAlphabet} \Pr(\msym |\mxst_\ell) \log_2 \Pr(\msym |\mxst_\ell)
  ~,
\label{eq:blackwellentropylimit}
\end{align}
where $\mxst_\ell = \mxst(w_{0:\ell})$ and $w_{0:\ell}$ is the first $\ell$
symbols of an arbitrarily long sequence $w_{0:\infty}$ generated by the
process. 

Note that $w_{0:\ell}$ will be a typical trajectory, if $\ell$ is sufficiently
long. To remove convergence-slowing contributions from transient mixed states,
one can ignore some number of the initial mixed states. The exact number of
transient states that should be ignored is unknown in general. That said, it
depends on the initial mixed state $\mxst_0$, which is generally taken to be
$\StartMS$, and the diameter of the attractor.

This completes our development of the HMC entropy rate. Appendix
\ref{app:Examples} applies the theory and associated algorithm to a number of
examples, with both countable and uncountable mixed states, and reveals 
a number of surprising properties. We now turn to practical issues of the
resources needed for accurate estimation.

\subsection{Data Requirements}
\label{sec:DataRequirements}

Although we developed our HMC entropy-rate expression in terms of IFSs,
determining a process' entropy rate can be recast as Markov chain Monte Carlo
(MCMC) estimation. In MCMC, the mean of a function $f(x)$ of interest over a
desired probability distribution $\pi(x)$ is estimated by designing a Markov
chain with a stationary distribution $\pi$. For HMCs the desired distribution is
the Blackwell measure $\MxSMeasure$, which is the stationary distribution $\mu$
over the MSP states $\MxSSet$. Then, the Markov chain is simply the transition
dynamic $\MxSDyn$ over $\MxSSet$.

With this setting, we estimate the entropy rate $\widehat{\hmu}^B$ as the mean
of the stochastic process defined by taking the entropy
$H[\MeasSymbol_{\mxst}]$ over symbols emitted from state $\mxst$ for a sequence
of mixed states generated by $\MxSDyn$. In effect, we estimate the entropy rate
as the mean of this stochastic process:
\begin{align}
\widehat{\hmu}^B & = \mu_{H} \nonumber \\
  & = \langle H[\MeasSymbol_{\mxst}]\rangle_{\mu}
  ~.
\label{eq:MarkovAverage}
\end{align}

Mathematically, little has changed. The advantage, though, of this alternative
description is that it invokes the extensive body of results on MCMC
estimation. In this, it is well known that there are two fundamental sources of
error in the estimation. First, there is that due to \emph{initialization bias}
or undesired statistical trends introduced by the initial transient data
produced by the Markov chain before it reaches the desired stationary
distribution.  Second, there are errors induced by \emph{autocorrelation in
equilibrium}. That is, the samples produced by the Markov chain are correlated.
And, the consequence is that statistical error cannot be estimated by $1 /
\sqrt{N}$, as done for $N$ independent samples.

To address these two sources of error, we follow common MCMC practice,
considering two ``time scales'' that arise during estimation. Consider the
autocorrelation of the stationary stochastic process:
\begin{align*}
C_f(t) = \langle f_s f_{s+t} \rangle - \mu_f^2
  ~,
\end{align*}
where $\mu_f$ is $f$'s mean. Also, consider the \emph{normalized}
autocorrelation, defined:
\begin{align*}
\rho_f(t) = \frac{C_f(t)}{C_f(0)}
 ~.
\end{align*}
If the autocorrelation decays exponentially with time, we define the
\emph{exponential autocorrelation time}:
\begin{align*}
\tau_{exp, f} = \lim_{t\to\infty} \sup \frac{1}{- \log | \rho_f(t) | }
\end{align*}
and
\begin{align*}
\tau_{exp} = \sup_{f} \tau_{exp, f}
  ~.
\end{align*}
So, $\tau_{exp}$ upper bounds the rate of convergence from an initial
nonequilibrium distribution to the equilibrium distribution.

For a given observable, we also define the \emph{integrated autocorrelation
time} $\tau_{init, f}$ as:
\begin{align}
\tau_{int, f} = \frac{1}{2} \sum_{-\infty}^{\infty} \rho_f(t)
  ~.
\end{align}
This relates the correlated samples selected by the chain to the variance of
independent samples for the particular function $f$ of interest. The variance of
$f(x)$'s sample mean in MCMC is higher by a factor of $2\tau_{int, f}$. In
other words, the errors for a sample of length $N$ are of order
$\sqrt{\tau_{int,f} / N}$. Thus, targeting $1\%$ accuracy requires $\approx
10^4 \tau_{int, f}$ samples. 

In practice, it is difficult to find $\tau_{exp}$ and $\tau_{int}$ for a
generic Markov chain. There are two options. The first is to use numerical
approximations that estimate the autocorrelation function, and therefore
$\tau$, from data. If we have the nonunifilar model in hand, it is a simple
matter of sweeping through increasingly long strings of generated data until
we observe convergence of the autocorrelation function.

Alternatively, taking inspiration from previous treatments of nonunifilar
models, we make a finite-state approximation to the MSP by coarse-graining the
simplex into boxes of length $\epsilon$ and employ a suitable method, such as
Ulam's method, to approximate the transition operator. Using methods previously
discussed in Ref. \cite{Riec17a}, this allows calculating the autocorrelation
function directly. \Cref{app:PredictiveDistortion} shows that that the
approximation error vanishes as $\epsilon \to 0$.

The net result is that, being cognizant of the data requirements, entropy rate
estimation is well behaved, convergent, and accurate.

\section{Conclusion}

We opened this development considering the role that determinism and randomness
play in the behavior of complex physical systems. A central challenge in this
has been quantifying randomness, patterns, and structure and doing so in a
mathematically-consistent but calculable manner. For well over a half a century
Shannon entropy rate has stood as the standard by which to quantify randomness
in a time series. Until now, however, calculating it for processes generated by
nonunifilar HMCs has been difficult, at best.

We began our analysis of this problem by recalling that, in general, hidden
Markov chains that are not unifilar have no closed-form expression for the
Shannon entropy rate of the processes they generate. Despite this, these HMCs
can be \emph{unifilarized} by calculating the mixed states. The resulting
mixed-state presentations are themselves HMCs that generate the process.
However, adopting a unifilar presentation comes at a heavy cost: Generically,
they are infinite state and so Shannon's expression cannot be used.
Nonetheless, we showed how to work constructively with these mixed-state
presentations. In particular, we showed that they fall into a common class of
dynamical system. The mixed-state presentation is an iterated function system.
Due to this, a number of results from dynamical systems theory can be applied.

Specifically, analyzing the IFS dynamics associated with a finite-state
nonunfilar HMC allows one to extract useful properties of the original process.
For instance, we can easily find the entropy rate of the generated process from
long orbits of the IFS. That is, one may select any arbitrary starting point in
the mixed-state simplex and calculate the entropy over the IFS's
place-dependent probability distribution. We evolve the mixed state according
to the IFS and sequentially sample the entropy of the place-dependent
probability distribution at each step. Using an arbitrarily long word and
taking the mean of these entropies, the method converges on the process'
entropy rate.

Although others consider the IFS-HMC connection \cite{rezaeian2006hidden,
Slomczynski_2000}, our development expanded previous work to include the much
broader, more general class of nonunifilar HMCs. In addition, we demonstrated
not only the mixed-state presentation's role in calculating the entropy rate,
but also its connection to existing approaches to randomness and structure in
complex systems. In particular, while our results focused on quantifying and
calculating a process' randomness, we left open questions of pattern and
structure. However, the path to achieving the results introduced here strongly
suggests that the mixed-state presentation offers insight into answering these
questions. For instance, \cref{Fig:simplextomachine} demonstrated how the
highly structured nature of the Simple Nonunifilar Source is made topologically
explicit through calculating its mixed-state presentation---which is also its
\eM. 

Though space will not let us develop it further here, this connection is not
spurious. Indeed, many information-theoretic properties of the underlying
process may be directly extracted from its mixed-state presentation. This
follows from our showing how the attractor of the IFS defined by an HMC is
exactly the set of mixed states $\MxSSet$ of that HMC. These sets are often
fractal in nature and quite visually striking. See
\cref{fig:sarahmachinesimplex} for several examples.

The sequel \cite{Jurg20c} to this development establishes that the fractal
dimension of the mixed-state attractor is exactly the divergence rate of the
\emph{statistical complexity} \cite{Crut88a}---a measure of a process'
structural complexity that tracks memory. Furthermore, the sequel introduces a
method to calculate the fractal dimension of the mixed-state attractor from the
Lyapunov spectrum of the mixed-state IFS. In this way, it demonstrates that
coarse-graining the simplex---the previous approach to study the structure of
infinite-state processes---may be avoided altogether.

To close, we note that these structural tools and the entropy-rate method
introduced here have already been put to practical use in two previous works.
One diagnosed the origin of randomness and structural complexity in quantum
measurement \cite{Vene19a}. The other exactly determined the thermodynamic
functioning of Maxwellian information engines \cite{Jurg20a}, when there had
been no previous method for this. At this point, however, we must leave the
full explication of these techniques and further analysis on how mixed states
reveal the underlying structure of processes generated by hidden Markov chains
to the sequel \cite{Jurg20c}. 

\section*{Acknowledgments}
\label{sec:acknowledgments}

The authors thank Sam Loomis, Greg Wimsatt, Ryan James, David Gier, and
Ariadna Venegas-Li for helpful discussions and the Telluride Science Research
Center for hospitality during visits and the participants of the Information
Engines Workshops there. JPC acknowledges the kind hospitality of the Santa Fe
Institute, Institute for Advanced Study at the University of Amsterdam, and
California Institute of Technology for their hospitality during visits. This
material is based upon work supported by, or in part by, FQXi Grant number
FQXi-RFP-IPW-1902, and U.S. Army Research Laboratory and the U.S. Army Research
Office under contract W911NF-13-1-0390 and grant W911NF-18-1-0028.

\appendix

\onecolumngrid
\clearpage
\begin{center}
{\huge Supplementary Materials}\\
\vspace{0.1in}
\vspace{0.1in}
{\huge The Shannon Entropy Rate of\\[10pt]
Hidden Markov Processes}\\[15pt]
{\large Alexandra Jurgens and James P. Crutchfield\\[5pt]
\arxiv{2002.XXXXX}
}
\end{center}

\setcounter{equation}{0}
\setcounter{figure}{0}
\setcounter{table}{0}
\setcounter{page}{1}
\setcounter{section}{0}
\makeatletter
\renewcommand{\theequation}{S\arabic{equation}}
\renewcommand{\thefigure}{S\arabic{figure}}
\renewcommand{\thetable}{S\arabic{table}}

The Supplementary Materials to follow review the notion of typical sets of
realizations in a stochastic process, discuss minimality of infinite-state
mixed-state presentations, determine the entropy rates of a suite of example
hidden Markov chains with infinite mixed-state presentations, and give details
of errors that arise when estimating autocorrelation.

\section{Asymptotic Equipartition and the Typical Set Contraction}
\label{app:AEP}

The \emph{asymptotic equipartition property} (AEP) states that for a discrete-time, ergodic, stationary process $X$:
\begin{align}
-\frac{1}{n} \log_2 \Pr(X_1, X_2, \dots, X_n) \to \hmu(X)
  ~,
\end{align}
as $n \to \infty$ \cite{Cove06a}. This effectively divides the set of
sequences into two sets: the \emph{typical set}---sequences for which the AEP
holds---and the atypical set, for which it does not. As a consequence of
the AEP, it must be the case that the typical set is measure one in the space
of all allowed realizations and all sequences in the atypical set approach
measure zero as $n \to \infty$. 

We argue that while our IFS class includes reducible maps, any composition of
maps corresponding to a word in the typical set will be irreducible. This can
be seen intuitively by considering the SNS, shown in \cref{fig:SNS}, and adding
an additional transition on a $\square$ from $\sigma_0$ to $\sigma_1$. This
produces an HMC with two reducible symbol-labeled transition matrices, but an
irreducible total transition matrix. However, as $|w| \to \infty$, the only
words such that $T^{(w)}$ remains reducible are $\square^N$ and $\triangle^N$.
We can see that these words cannot possibly be in the typical set, since
$-\frac{1}{n} \log_2 \Pr(\square^n) = -\log_2\Pr(\square) \neq \hmu(X)$. The
entropy rate $\hmu$ is by definition the branching entropy averaged over the
mixed states. And so, any word that visits only a restricted subset of the
mixed states---i.e., a word with a reducible transition matrix---cannot
approach $\hmu$, regardless of length. Therefore, only words with an
irreducible mapping will be in the typical set, implying that there exists an
integer word length $|w| > 0$ for which words without a contractive mapping are
measure zero.

\section{Minimality of $\MSP(M)$}
\label{app:MinimalityofMSP}

The minimality of infinite-state mixed-state presentations $\MSP(M)$ is an
open question. As demonstrated in \cref{app:cantorsetmachine}, it is possible to
construct MSPs with an uncountably infinite number of states for a process that
requires only one state.

A proposed solution to this problem is a short and simple check on
\emph{mergeablility} of mixed states, which here refers to any two distinct
mixed states that have the same conditional probability distribution over
future strings; i.e., any two mixed states $\mxst_0$ and $\mxstalt_0$ for
which:
\begin{align}
\Pr( \MS{0}{L} | \mxst_0 ) = \Pr( \MS{0}{L} | \mxstalt_0 )
  ~,
\label{eq:MergeableStates}
\end{align}
for all $L \in \mathbb{N}^+$.

\begin{figure}[h]
\centering
\includegraphics{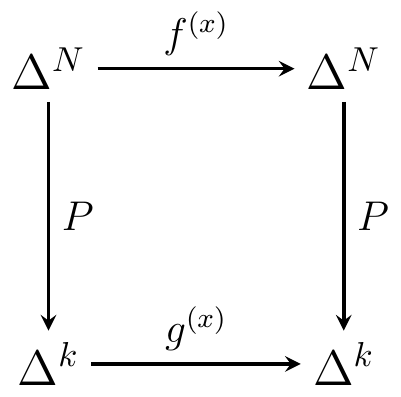}
\caption[text]{Commuting diagram for probability functions $P = \{p^{(x)} \}$,
	mixed-state mapping functions $f^{(x)}$, and proposed
	symbol-distribution mapping functions $g^{(x)}$.
	}
\label{Fig:simplexdiagram}
\end{figure}

Although minimality does not impact the entropy-rate calculation, one benefit
of the IFS formalization of the MSP is the ability to directly check for
duplicated states and therefore determine if the MSP is nonminimal. We check
this by considering, for an $N+1$ state machine $M$ with alphabet
$\MeasAlphabet = \{ 0, 1, \dots, k \}$, the dynamic not only over mixed states,
but probability distributions over symbols. Let:
\begin{align}
P(\mxst) = \left( p^{(0)}(\mxst), \dots, p^{(k-1)}(\mxst) \right)
\label{eq:VectorProbabilityFunction}
\end{align}
and consider \cref{Fig:simplexdiagram}. For each mixed state $\mxst \in
\simplex^N$, \cref{eq:VectorProbabilityFunction} gives the corresponding
probability distribution $\rho(\mxst) \in \simplex^k$ over the symbols
$\msym \in \MeasAlphabet$. Let $M$ emit symbol $\msym$, then the dynamic from one such probability distribution $\rho \in \simplex^{k}$ to the next
is given by:
\begin{align}
g^{(\msym)}(\rho_t) & = P \circ f^{(\msym)} \circ P^{-1} (\rho) \nonumber \\
  & = \rho_{t+1, \msym}
  ~.
\label{eq:ProabilityVectorDynamic}
\end{align}

From this, we see that if \cref{eq:VectorProbabilityFunction} is invertible,
$g^{(\msym)} : \simplex^k \to \simplex^k$ is well defined and has the same
functional properties as $f^{(\msym)}$. In other words, in this case, it is not
possible to have two distinct mixed states $\mxst, \mxstalt \in \simplex^N$
with the same probability distribution over symbols. And, the probability
distributions can only converge under the action of $g^{(\msym)}$ if the mixed
states also converge under the action of $f^{(\msym)}$. Shortly, we consider
several cases where $P$ is \emph{not} invertible over the entire symbol simplex.

If every mixed state in $\MxSSet$ corresponds to a unique probability
distribution over symbols, we conjecture that the corresponding $\MSP(M)$ is the
minimal unifilar representation of the underlying process $\Process$. If we
then trim the transient states of $\MSP(M)$, leaving the recurrent set
$\MxSSet_{\text{R}}$, the result is the \eM.

\section{Examples} 
\label{app:Examples}

The following illustrates how to apply the theory and algorithms from the main
text to accurately and efficiently calculate the entropy rate of processes
generated by HMCs with countable and uncountable mixed states. It highlights a
number of curious and nontrivial properties of these processes and their MSPs.

\subsection{Cantor Set MSP}
\label{app:cantorsetmachine}

We first analyze a process with an MSP whose uncountable mixed states lie in a
Cantor set. Surprisingly, this MSP is far from minimal, as the process is, in
fact, generated by a biased coin---that is, a single-state \eM.

\subsubsection{The Cantor Set}
\label{app:thecantorset}

The Cantor set is perhaps the most well-known example of a nontrivial
self-similar (fractal) set. The familiar middle-thirds version is constructed by
starting with the unit interval $C_1 = [0,1]$ and removing the middle third,
giving the set $C_2 = [0,\frac{1}{3}]\cup[\frac{2}{3}, 1]$. Repeating this with
each remaining subinterval in $C_2$ produces $C_3$, and so on. The Cantor set $C$ consists of points that remain after infinitely repeating this action:
$\text{C} = \bigcap_{n=1}^{\infty} C_n$.

The Cantor set is uncountably infinite and has Hausdorff dimension:
\begin{align*}
\dim_H \text{C} = \frac{\log 2}{\log 3}
  ~.
\end{align*}
A parametrized family of Cantor sets is generated by repeating $C_2 =
[0,\frac{1}{s}]\cup[\frac{s-1}{s}, 1]$ (i.e., removing the middle
$\frac{s-2}{s}$), the Hausdorff dimension is:
\begin{align*}
\dim_H \text{C} = \frac{\log 2}{\log s}
  ~.
\end{align*}
Simply stated, the dimension is the logarithm of the number of copies of the
original unit interval made at each iteration, divided by the logarithm of the
length ratio between the original object and its copy.

\subsubsection{The Cantor Machine}
\label{app:thecantormachine}

\begin{figure}
\centering
\includegraphics{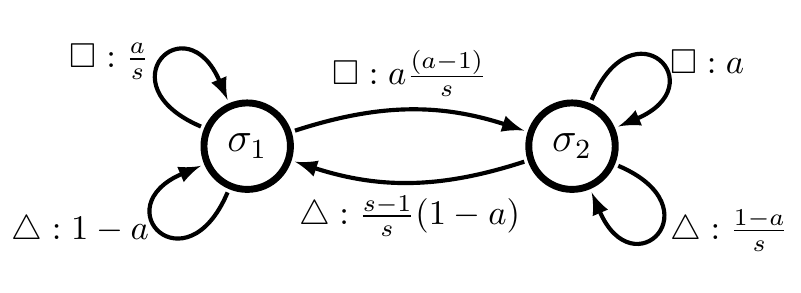}
\caption[text]{Nonunifilar HMC $M_C$ that generates Cantor sets of mixed states
	on the $2$-simplex, for values of $0 < p < 1$ and $ s > 2$. For $s=3$, we
	find the middle-third Cantor set.
	}
\label{Fig:cantormachine}
\end{figure}

The Cantor set, due to its familiarity, makes for a useful, first object of
study for uncountable-mixed-state HMCs. \Cref{Fig:cantormachine} shows an
HMC $M_{\text{C}}$ that generates a Cantor set of mixed states. There $0 < a <
1$ adjusts the statistical bias of the measure over the Cantor set and $s > 2$
is the scaling ratio for copying the intervals.

From \cref{Fig:cantormachine} we read off the transition matrices:
\begin{align*}
T^{(\square)} = \begin{pmatrix}
     \frac{a}{s}  & \frac{a(s-1)}{s}  \\
     0 & a
 \end{pmatrix}
~\text{and}~
T^{(\triangle)} \begin{pmatrix}
     1-a & 0  \\
     \frac{(s-1)(1-a)}{s} & \frac{1-a}{s}
\end{pmatrix}
  ~.
\end{align*}
This allows us to immediately write down the probability functions and
mapping functions, recalling that in the two-state case the vectors on the
simplex take the form $\bra{\mxst} = (\langle \mxst \ket{\delta_1}, 
1- \langle \mxst \ket{\delta_1})$:
\begin{align*}
\left\{ \begin{aligned}
  & p^{(\square)} ( \mxst )= \bra{ \mxst } T^{(\square)} \ket{\One} = a \\
  & p^{(\triangle)} ( \mxst ) = \bra{ \mxst } T^{(\triangle)} \ket{\One} = 1-a 
  \end{aligned}
\right.
\end{align*}
and:
\begin{align*}
\left\{ \begin{aligned}
         & f^{(\square)} ( \mxst ) = \frac{ \bra{ \mxst } 
         T^{(\square)} }{\bra{ \mxst } T^{(\square)} \ket{\One}} 
         = \left(\frac{ \langle \mxst \ket{\delta_1} }{s}, 1 - 
         \frac{\langle \mxst \ket{\delta_1}}{s} \right) \\
         & f^{(\triangle)} ( \mxst ) = \frac{ \bra{ \mxst } 
         T^{(\triangle)} }{\bra{ \mxst } T^{(\triangle)} \ket{\One}} 
         = \left(\frac{s +\langle \mxst \ket{\delta_1} - 1}{s},
          \frac{1 -\langle \mxst \ket{\delta_1}}{s}\right)
         \end{aligned}
\right.
  ~.
\end{align*}
It is easily seen, by considering $\langle \mxst \ket{\delta_1} = 0$ and
$\langle \mxst \ket{\delta_1} = 1$, that these maps, in fact, map the simplex
to the first and second intervals of $C_2$, respectively.

\begin{figure}\centering
\includegraphics[width=0.5\textwidth]{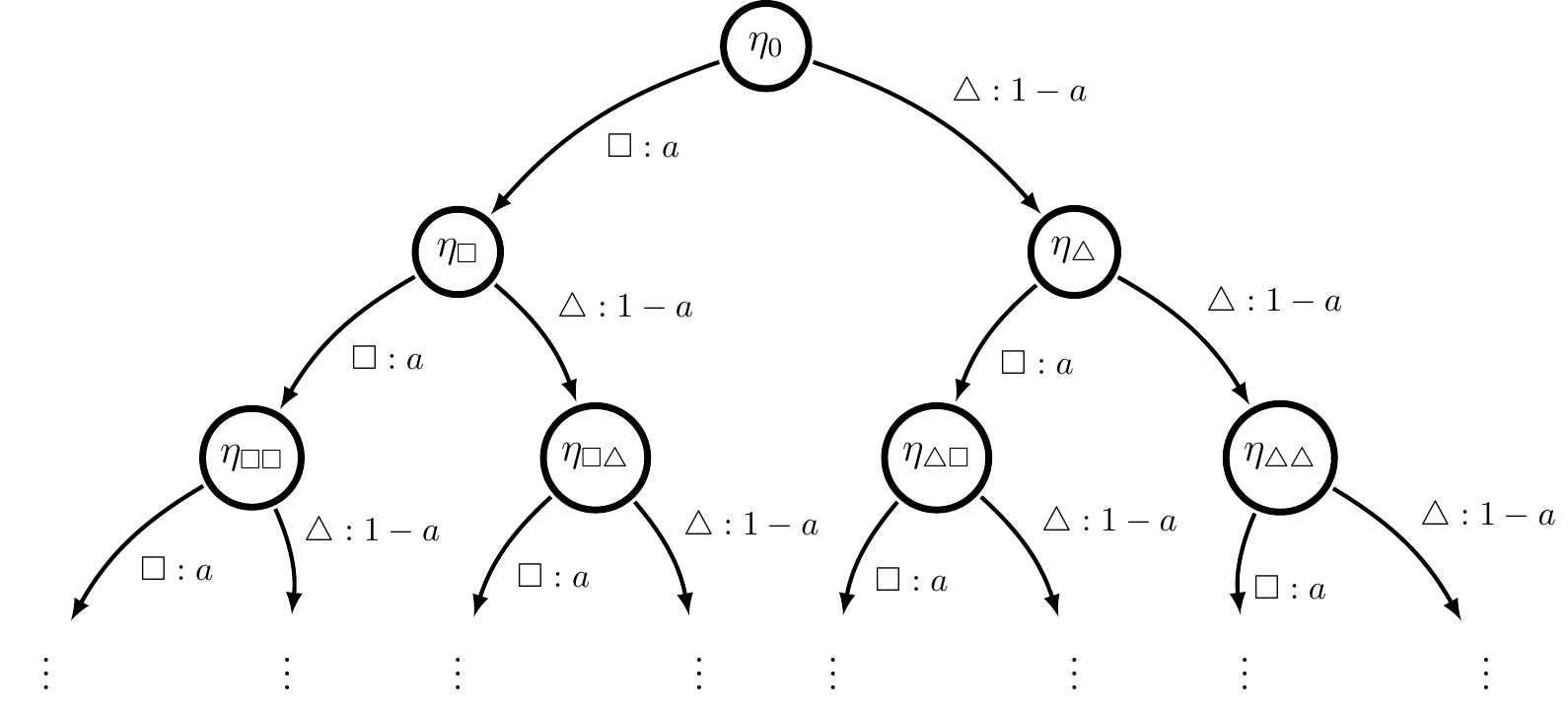}
	\\[20pt]
\includegraphics{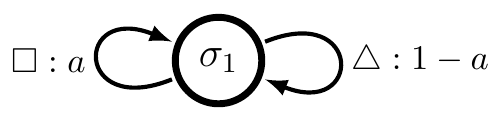}
\caption{Two valid alternative presentations of the Cantor set machine:
	(Top) The MSP $\MSP(M_{\text{C}})$ of the Cantor set machine $M_{\text{C}}$
	in \cref{Fig:cantormachine}. The set of mixed states $\eta_w$ is
	uncountably infinite. (Bottom) A unifilar hidden Markov model, commonly
	called the Biased Coin, that generates the same process as the nonunifilar
	Cantor machine $M_{\text{C}}$ in \cref{Fig:cantormachine}.
	}
\label{fig:cantormachinemixedstates}\end{figure}

The Cantor Machine MSP $\MSP(M_{\text{C}})$ is shown
\cref{fig:cantormachinemixedstates} (Top). It has an uncountably-infinite
number of recurrent states, which correspond exactly to the elements of the
Cantor set. Since the probability functions do not depend on $\bra{ \mxst }$,
we do not need to invoke the Ergodic Theorem, but instead can calculate the
entropy exactly: 
\begin{align*}
h_{ \MxSMeasure_{\MSP(M_{\text{C}})} } & = - \int \sum_i p^{(\msym)} 
  ( \mxst ) \log_2  p^{(\msym)} (\mxst )
  d \MxSMeasure_{\MSP(M_{\text{C}})} (\mxst) \\
  &  = - a \log_2 (a) - (1 - a) \log_2 (1 - a) \\
  & = H \left( a \right) ~.
\end{align*}

\subsubsection{A Biased Coin}
\label{app:abiasedcoin}

However, there is an important caveat here, noted in
\cref{app:MinimalityofMSP}. The MSP may contain states that are
probabilistically equivalent. The probability mapping functions are
noninvertible and, in fact, every single mixed state corresponds to the same
conditional probability distribution over symbols. This means that the
uncountably-infinite MSP is not a minimal presentation. There is a markedly
simpler unifilar model for the Cantor set machine $M_{\text{C}}$.

In fact, all mixed states in $\MSP(M_{\text{C}})$ collapse into a single state,
giving the minimal unifilar model of the Cantor set machine as the Biased Coin
HMC shown in \cref{fig:cantormachinemixedstates} (Bottom). This HMC generates
the same process as the Cantor machine, but requires only a single state.

\begin{figure}[b]
\centering
\includegraphics{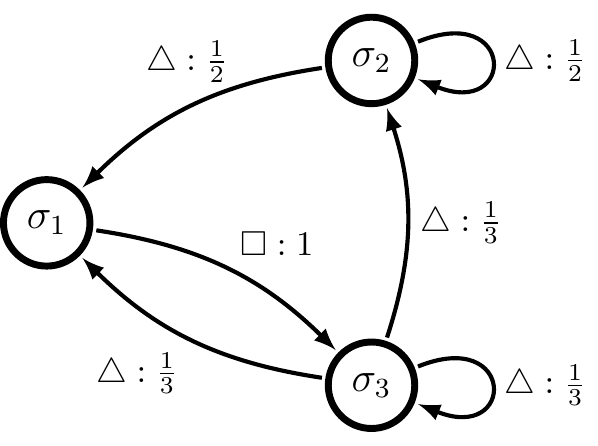}
\caption[text]{Three-state, nonunifilar machine $M_{3S}$.}
\label{Fig:threestate}
\end{figure}

\subsection{Countable MSP with $2$-state HMC}
\label{sec:threestatemachine}

Now, we explore a different, but related case that introduces a condition for
a countable MSP and again highlights the role of minimality.

\subsubsection{$3$-State HMC with a Countable MSP}

Consider the $3$-state HMC $M_{3S}$ of \cref{Fig:threestate}. The transition
matrices for this machine are:
\begin{align*}
T^{(\square)} = \begin{pmatrix}
     0 & 0 & 1 \\
     0 & 0 & 0 \\
     0 & 0 & 0
 \end{pmatrix}
 ~\text{and}~
T^{(\triangle)} = \begin{pmatrix}
     0 & 0 & 0  \\
     \frac{1}{2} & \frac{1}{2} & 0 \\
     \frac{1}{3} & \frac{1}{3} & \frac{1}{3}
 \end{pmatrix}
  ~.
\end{align*}
These give the mapping and probability functions:
\begin{align*}
 \left\{ \begin{aligned}
         & p^{(\square)} (\mxst)= \mxst_1 \\
         & p^{(\triangle)} (\mxst) = 1 - \mxst_1 
         \end{aligned}
 \right.
\end{align*}
and:
\begin{align*}
 \left\{ \begin{aligned}
         & f^{(\square)} (\mxst) = \left( 0, 0, 1 \right) \\
         & f^{(\triangle)} (\mxst) = \left( \frac{1}{2}\mxst_2 + \frac{1}{3}\mxst_3, \frac{1}{2}\mxst_2 + \frac{1}{3}\mxst_3, \frac{1}{3}\mxst_3 \right)
         \end{aligned}
 \right.
  ~.
\end{align*}

Consider the probability functions first. $P$ is not invertible over all of
$\simplex^2$, but is partially invertible over a restricted domain. Given a
line in the simplex where $\mxst_2$ and $\mxst_3$ are a function of
$\mxst_1$---say, $(\mxst_1, \frac{1-\mxst_1}{2}, \frac{1-\mxst_1}{2})$---we
can invert $P(\mxst)$. The question becomes: What is the appropriate
restricted domain?

Note that for both $f^{(\square)}$ and $f^{(\triangle)}$, $\mxst_1 =
\mxst_2$. In the simplex this corresponds to all the mixed states lying along
a line in $\simplex^2$---the line $(\mxst_1, \mxst_1, 1-2\mxst_1)$. This,
then, is the restricted domain over which the states $\MSP(M_{3S})$
correspond to unique probability distributions. The fact that this space is a
line implies that the generative machine can be written with only two states.

The constancy of the mapping function for $\square$ contributes further
structure, ensuring that the set of mixed states will be countably infinite.
We can write the mixed states down in series, in terms of how many
$\triangle$s we have seen since the last $\square$:
\begin{align*}
 \mxst(\triangle^n) = \left( \frac{2(3^n - 2^n)}{4 \cdot 3^n - 3 \cdot 2^n} ,  \frac{2(3^n - 2^n)}{4 \cdot 3^n - 3 \cdot 2^n}, \frac{2^n}{4 \cdot 3^n - 3 \cdot 2^n} \right) 
  ~,
\end{align*}
where $n = 0$ is taken to be the mixed state $\mxst(\triangle^0) = (0, 0,
1)$. The transition probabilities for these states are:
\begin{align*}
  \left\{ \begin{aligned}
     \Pr\left(\square | \eta(\triangle^n) \right) = \frac{ 2(3^n - 2^n) }{4 \cdot 3^n - 3 \cdot 2^n} \\
     \Pr\left(\triangle | \eta(\triangle^n) \right) = \frac{ 2\cdot3^n - 2^n}{4 \cdot 3^n - 3 \cdot 2^n} 
         \end{aligned}
 \right.
  ~.
\end{align*}

\begin{figure}[h]
\centering
\includegraphics[width=0.8\textwidth]{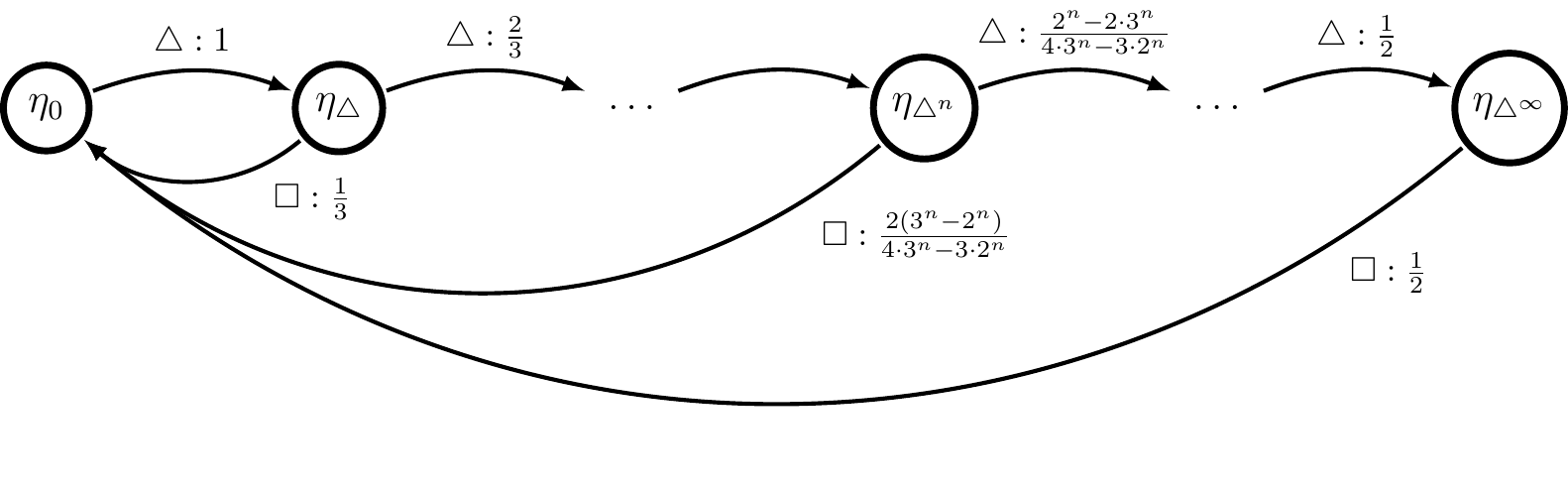}
\caption[text]{Unifilar HMC that generates the same process generated by the
	nonunfilar machine $M_{3S}$ in \cref{Fig:threestate}.
	}
\label{Fig:threestateuni}
\end{figure}

The MSP $\MSP(M_{3S})$ is shown in \cref{Fig:threestateuni}. If the initial
condition $\mxst_0 = (0, 0, 1)$, all mixed states generated by the mapping
functions are recurrent and, as we discussed, have unique probability
distributions. Therefore, the HMC in \cref{Fig:threestateuni} is the process'
\eM. Since $\MxSSet$ is countable, we can find $\MxSMeasure$ by hand, by
solving the set of equations $\pi_{n+1} = \pi_n \Pr(\triangle | \mxst_n)$,
with $\sum_n \pi_n = 1$. This gives $\pi_n = \frac{2}{5}\left( 2^{1-n} - 3^{-n}
\right)$ and we find: 
\begin{align*}
h_{\MxSMeasure_{\MSP(M_{3S})}} & = \sum_{n=0}^\infty
  \pi_n \log \Pr(\triangle | \mxst_n) \\
  & \approx 0.4381
 ~.
\end{align*}

\subsubsection{Actually, A 2-State Machine}
\label{sec:twostatemachine}

As mentioned above, the restricted domain over which $P$ is invertible implies
a smaller state set for the process generated by the nonunifilar machine
$M_{3S}$. For all relevant mixed states, $\Pr(\causalstate_1) =
\Pr(\causalstate_2)$, suggesting that we devise an HMC combining the two
states. However, the mapping function for $\square$ must still project
definitively to a single state, to retain the countable infinity of mixed
states. In fact, these restrictions ensure that the minimal nonunifilar HMC
for the process is the HMC for the Simple Nonunifilar Source, discussed in
\cref{sec:MixedStateasanIFS}.

If we declare that $\hat{\mxst}(\triangle^0) = (0,1)$, we may calculate the
subsequent sequence of mixed states associated with emitting an increasingly
long sequence of $\triangle$s, by using the mapping functions in
\cref{sec:MixedStateasanIFS}. The next two states are:
\begin{align*}
     \hat{\mxst}(\triangle^1) & = \left( 1-q, q \right) ~\text{and} \\
     \hat{\mxst}(\triangle^2) & = \left( \frac{1-p+pq-q^2}{1-p+pq}, 
     \frac{q^2}{1-p+ pq} \right)
  ~.
\end{align*}
For the underlying process to remain the same, the condition that must be met
is $P(\mxst(\triangle^n)) = \hat{P} ( \hat{\mxst} (\triangle^n))$. This
determines $p$ and $q$. For $n=0$ this is trivially met. For $n=1$ we have:
\begin{align*}
\hat{P} \left( \hat{\mxst} (\triangle^1) \right) 
  & = \big( p(1-q), 1 - (1-q)p \big) \\
  & = \left(\frac{1}{3}, \frac{2}{3} \right)
  ~,
\end{align*}
so that $1-q = \frac{p}{3}$. Substituting this into the
$\hat{\mxst}(\triangle^2)$ condition we get:
\begin{align*}
\hat{\mxst}(\triangle^2)
  = \left( 1 - \frac{3}{2} q^2, \frac{3}{2} q^2 \right)
  ~. 
\end{align*}

Substituting this into the probability distribution constraint for $n=2$ gives
$q = 1 / 2$ or $q = 1 / 3$, corresponding to two different $2$-state
nonunifilar HMCs that generate the same process as the $3$-state HMC. This
further emphasizes the lack of uniqueness of generative models. That said, by
examining the underlying IFS, their HMCs can be recovered.

\subsection{Parametrized HMCs and Their MSPs}
\label{sec:parametrizedmachines}

Finally, consider an HMC with $3$ symbols and $3$ states:
\begin{align}
 T^{\square} = \begin{pmatrix}
     \alpha y & \beta x & \beta x \\
     \alpha x & \beta y & \beta x \\
     \alpha x & \beta x & \beta y
 \end{pmatrix}
 ,
 \quad
 T^{\triangle} = \begin{pmatrix}
     \beta y & \alpha x & \beta x \\
     \beta x & \alpha y & \beta x \\
     \beta x & \alpha x & \beta y        
 \end{pmatrix}
 ,~\text{and}~
 \quad
 T^{\large\circ} = \begin{pmatrix}
     \beta y & \beta x & \alpha x \\
     \beta x & \beta y & \alpha x \\
     \beta x & \beta x & \alpha y        
 \end{pmatrix}
  ~,
\label{eq:sarah_machine}
\end{align}
with $\beta = \frac{1-\alpha}{2}$ and $y = 1 - 2x$. From inspection, we see
that $\alpha$ can take on any value from $0$ to $1$ and $x$ may range from
$0$ to $\frac{1}{2}$. 

\begin{figure}
 \centering
 \subfloat[100,000 mixed states of the HMC defined by 
 \cref{eq:sarah_machine} with $\alpha = 0.6$ and $x = 0.025$.]{
   \includegraphics[width=0.45\textwidth]{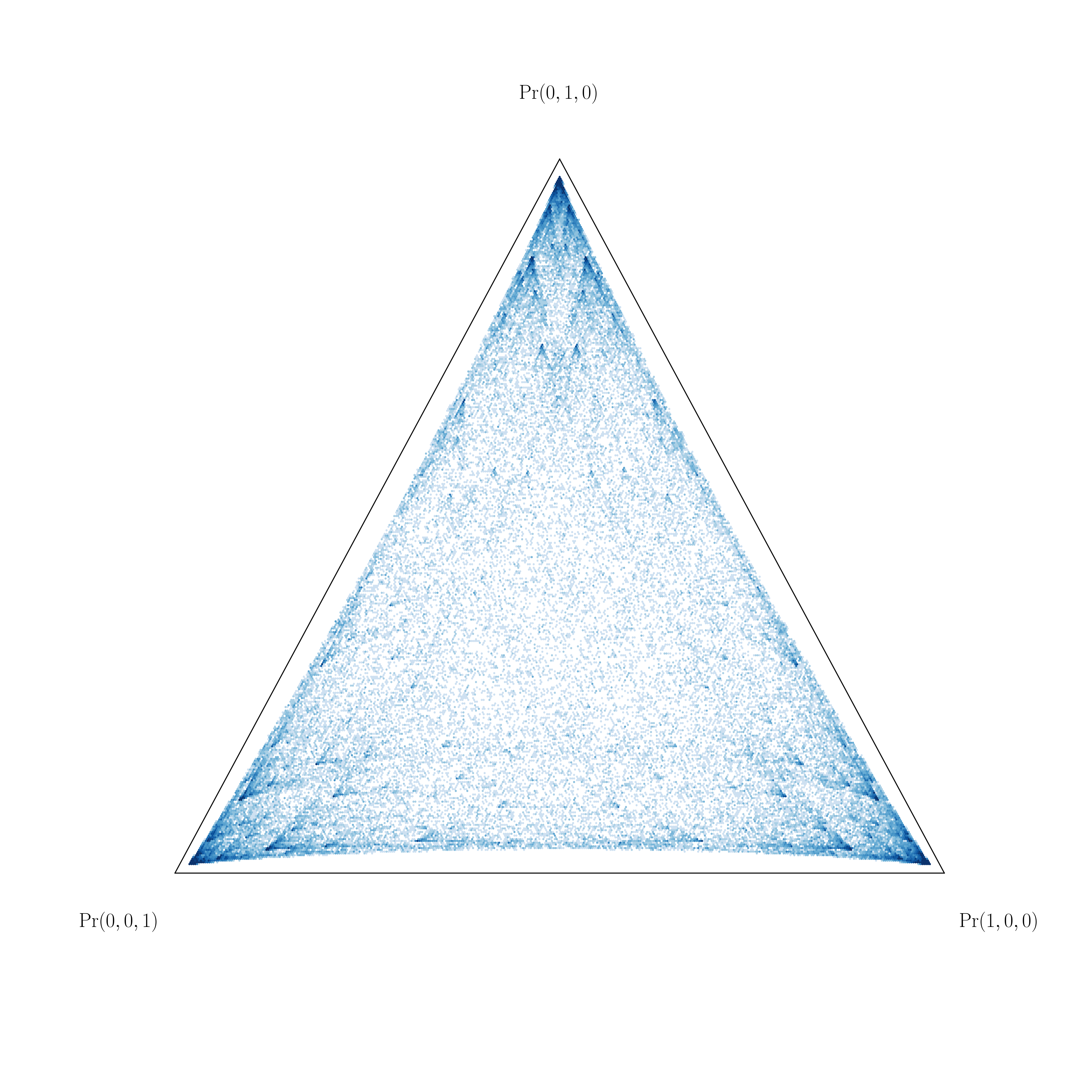}
 }
 \subfloat[100,000 mixed states of the HMC defined by 
 \cref{eq:sarah_machine} with $\alpha = 0.6$ and $x = 0.10$.]{
   \includegraphics[width=0.45\textwidth]{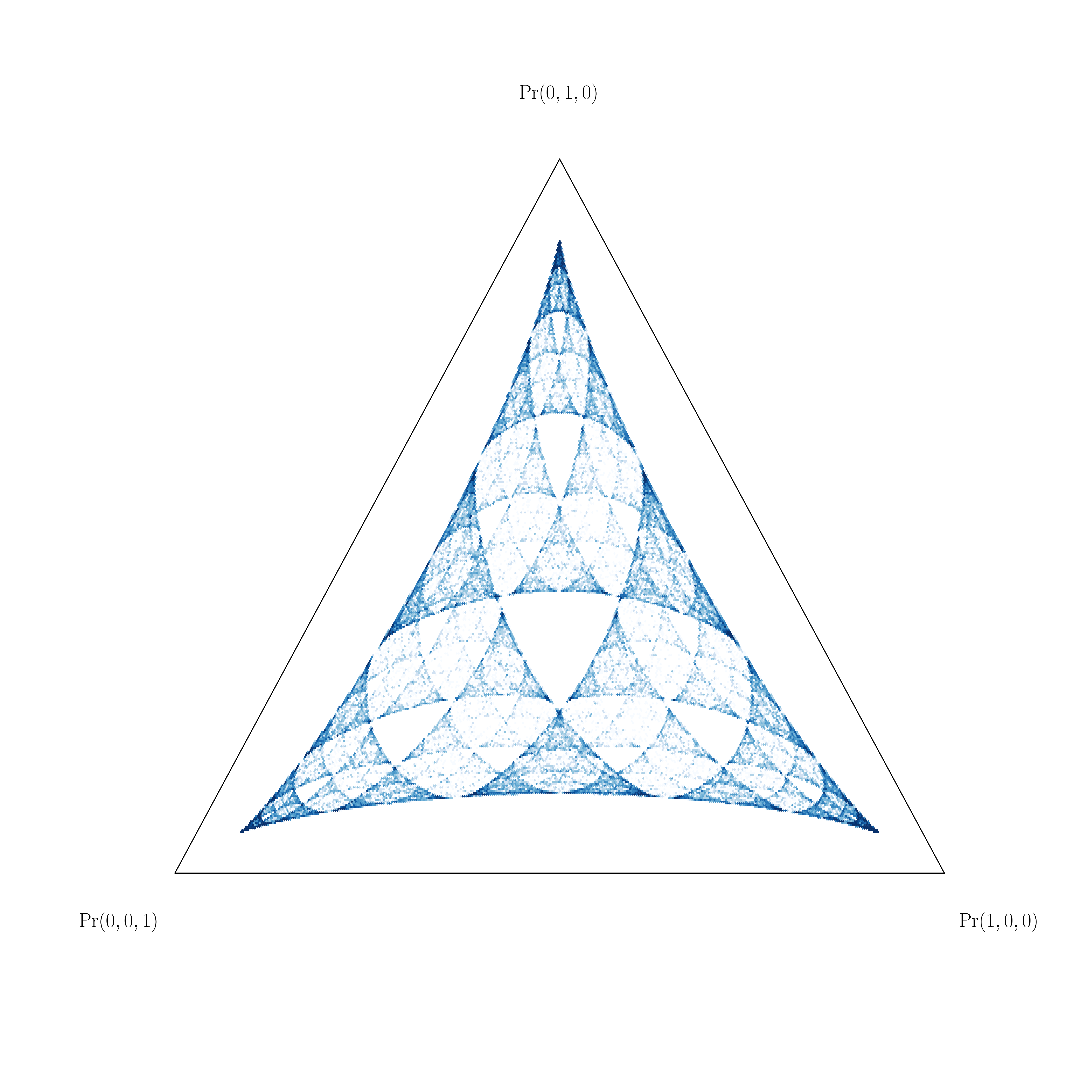}
 }
 \hspace{0mm}
 \subfloat[100,000 mixed states of the HMC defined by 
 \cref{eq:sarah_machine} with $\alpha = 0.6$ and $x = 0.33$.]{
   \includegraphics[width=0.45\textwidth]{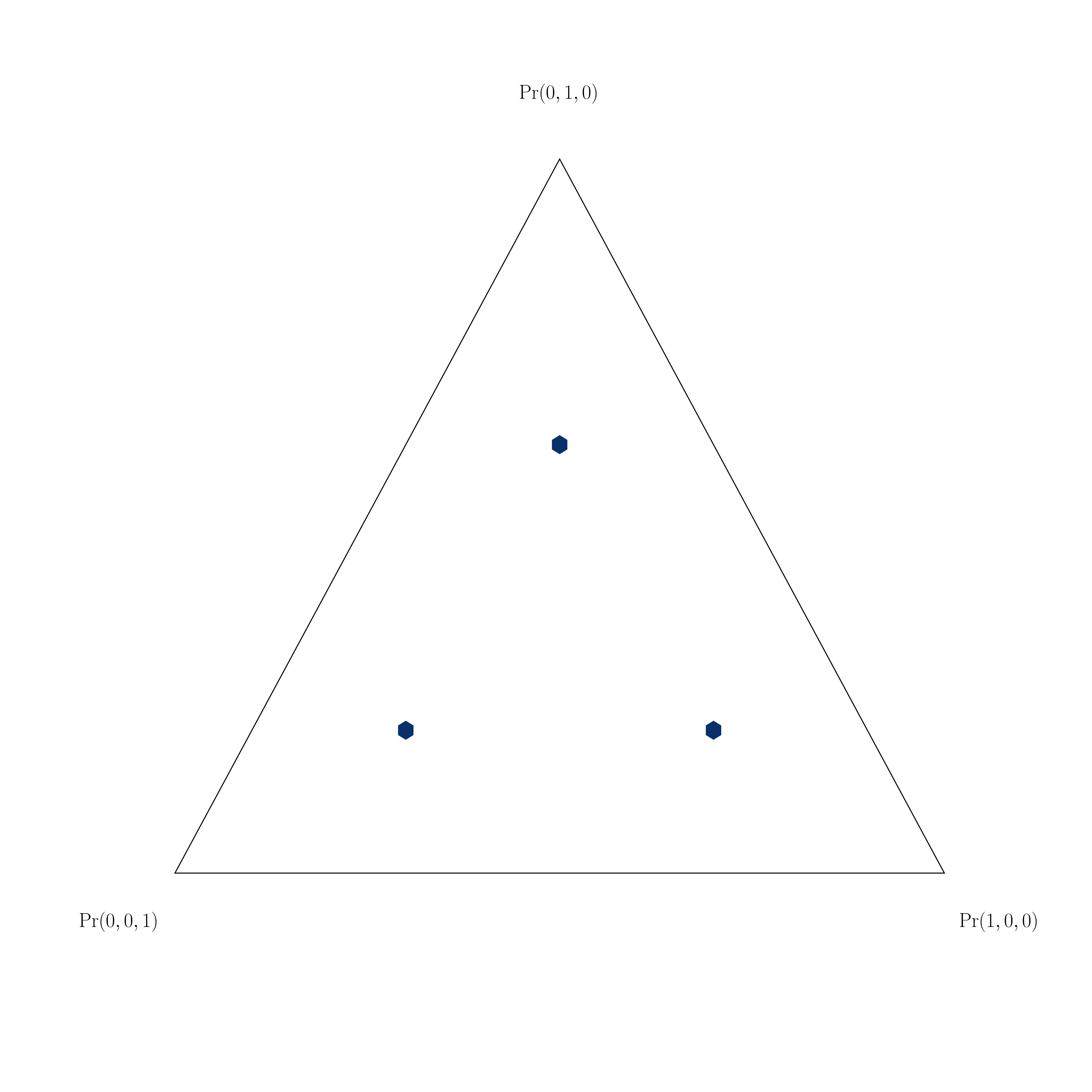}
 }
 \subfloat[100,000 mixed states of the HMC defined by 
 \cref{eq:sarah_machine} with $\alpha = 0.6$ and $x = 0.49$.]{
   \includegraphics[width=0.45\textwidth]{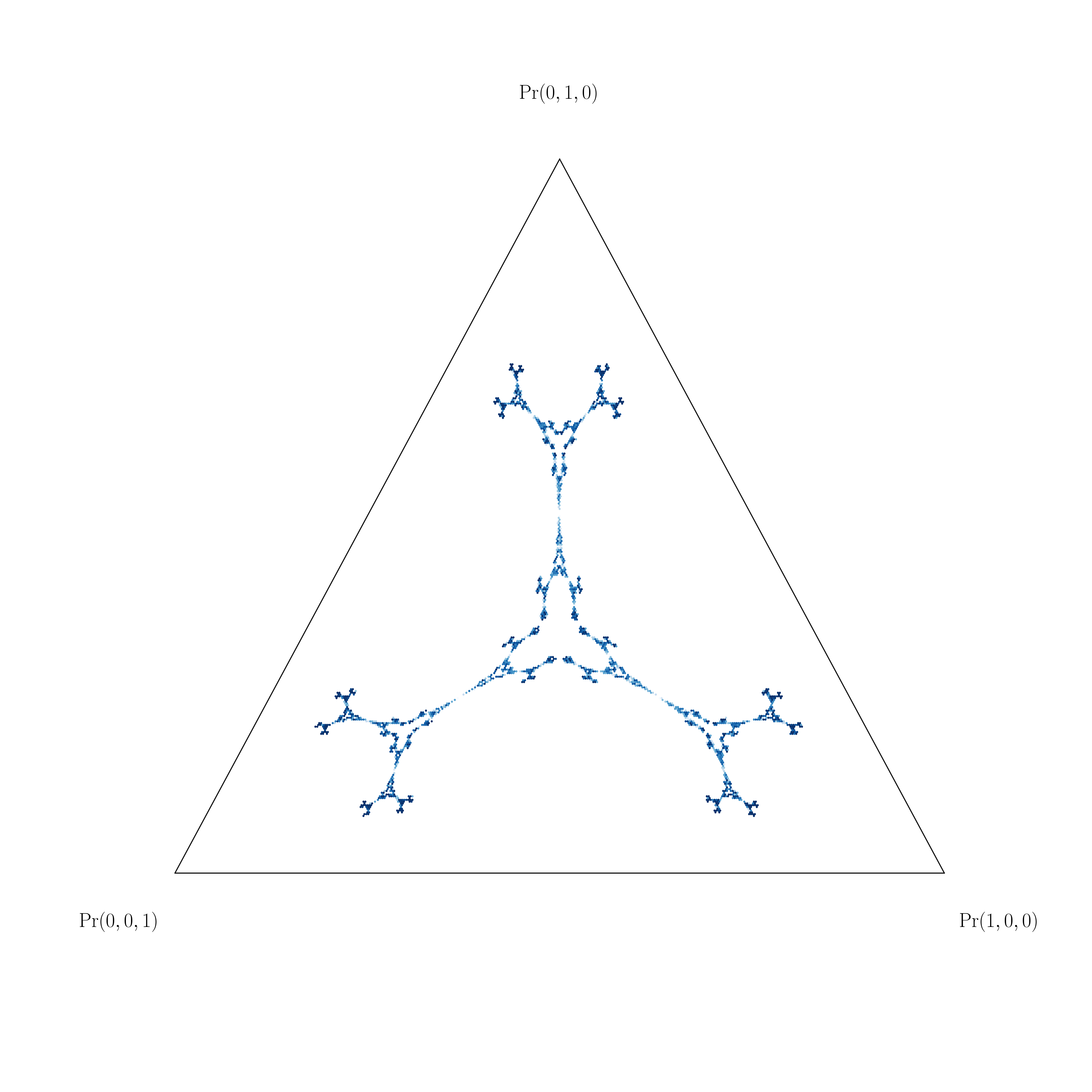}
 } 
\caption{Parametrized $3$-state HMC defined in \cref{eq:sarah_machine}
	that generates MSPs in a variety of structures, depending on $x$ and
	$\alpha$. However, due to the rotational symmetry in the transition
	matrices, the attractor is radially symmetric around the simplex center.
	}
\label{fig:sarahmachinesimplex}
\end{figure}

\begin{figure}
\centering
\includegraphics[width=0.8\textwidth]{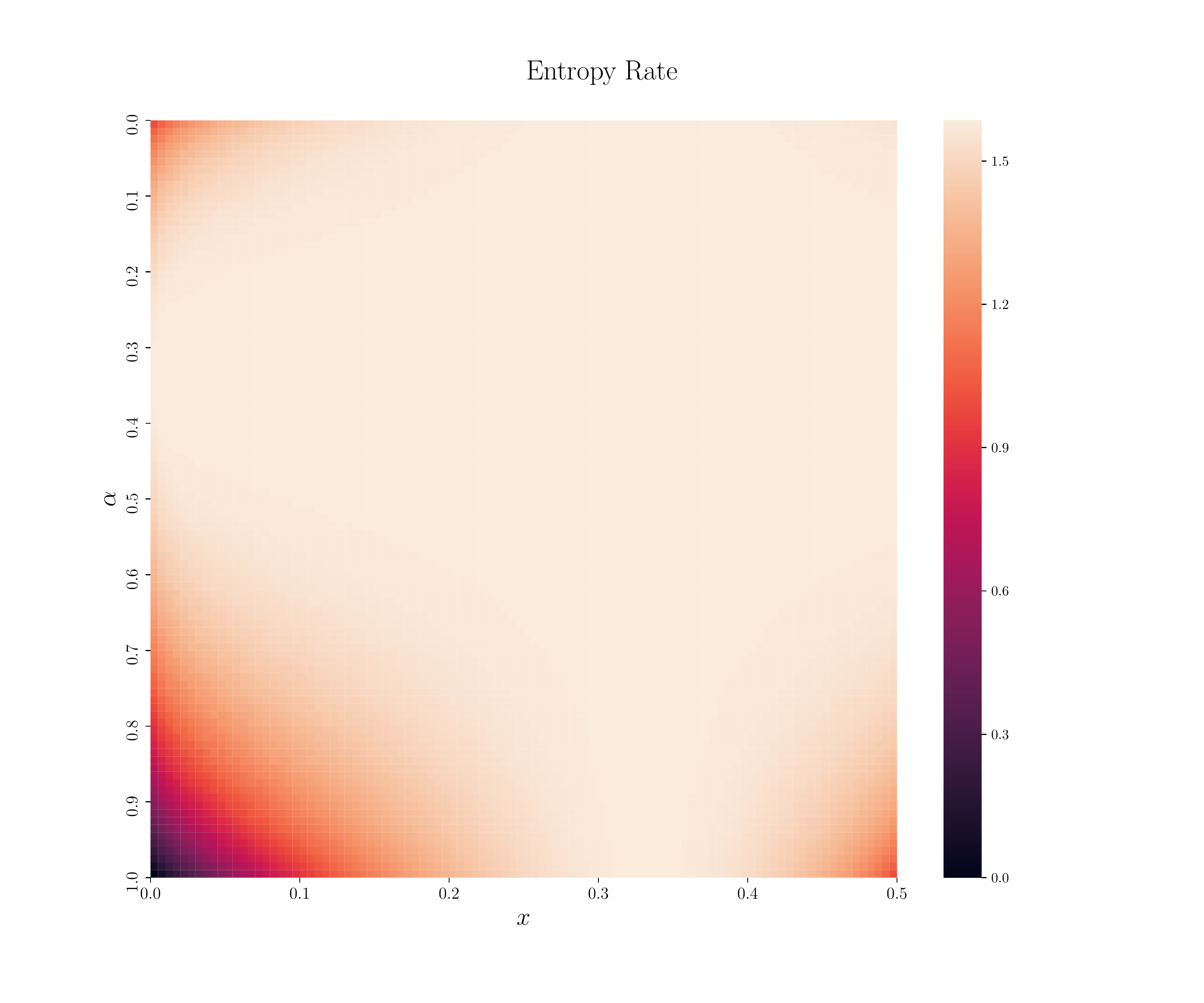}
\caption[text]{Entropy rates of the parametrized HMC defined in
	\cref{eq:sarah_machine} over $x \in [0.0,0.5]$ and $\alpha \in [0.0,1.0]$.
	}
\label{Fig:sarah07_entropy}
\end{figure}

Choosing $\alpha = 0.6$ and sweeping $x \in [0,0.5]$ gives us an MSP that first
fills nearly the entire simplex, with probability mass concentrated at the
corners, then shrinks to a finite machine with $3$ states at $x= 1 / 3$, and
finally grows once again into a fractal measure, as
\cref{fig:sarahmachinesimplex} illustrates. To demonstrate the ease and
efficiency calculating their entropy rates \cref{Fig:sarah07_entropy} plots
$\hmu$ as function of $(x,\alpha) \in [0,0.5] \times  [0,1]$. It is an
interesting side note that despite the wildly different structures on display in
\cref{fig:sarahmachinesimplex}, we see a smoothly varying entropy rate that does
not appear to be strongly affected by the underlying structure. This case and
the expected impact of structure on the entropy rate more broadly will be
discussed in further detail in the sequel. 

\section{Estimation Errors for Finite-State Autocorrelation}
\label{app:PredictiveDistortion}

Coarse-graining the mixed-state simplex into a set $\CoarseGraining$ of boxes of
width $\epsilon$, we may construct a finite-state approximation of the
infinite-state MSP. It has been shown that given such an approximation, for
any given box $c$, the bound on the difference in the entropy rate over the
symbol distribution between the coarse-grained approximation and a mixed
state within that box is bounded by:
\begin{align}
 \left| H[X_0 | \CoarseGraining = c] - H[X_0 | \eta \in f] \right| 
 \leq H_b \left(\frac{\sqrt{G}\epsilon}{2} \right),
 \label{eg:EntropyBound}
\end{align}
where $H_b(\cdot)$ is the binary entropy function \cite{Marz17a}. Our task
here is to consider the error in the autocorrelation in the sequence of
mixed states since, if we can show that this is bounded, the error in the
autocorrelation of the branching entropy must also be bounded. 

At time zero, the autocorrelation is equal to $A(L=0) = \langle X_0
\overline{X_0} \rangle$, so for the finite-state approximation, we have:
\begin{align*}
 A_{\CoarseGraining} (L = 0) = \sum_i \pi_{\CoarseGraining} (i) 
 c_i \overline{c_i}
  ~,
\end{align*}
where $\pi_{\CoarseGraining}$ is the stationary distribution over the
coarse-grained mixed states, $\pi_{\CoarseGraining}(i)$ is the stationary
probability of cell $i$, and $c_i$ is the center of cell $i$. For the true
process, we have:
\begin{align*}
A(L=0) & = \int_{\MxSSet} d\MxSMeasure(\mxst) \mxst \overline{\mxst} \\
  & = \sum_i \pi_{\CoarseGraining} (i)
  \int_{\mxst \in \CoarseGraining_i}
  d \MxSMeasure(\mxst | i) \mxst \overline{\mxst} 
  ~,
\end{align*}
where $d \MxSMeasure(\mxst | i)$ is the distribution over mixed states
within cell $i$. The maximum distance between any two mixed states in a cell
$i$ is bounded by:
\begin{align*}
\| \mxst - \mxstalt \|_1 \leq \sqrt{G} \epsilon
  ~,
\end{align*} 
the length of the longest diagonal in a hypercube of dimension $|G|$, by
construction. Since the gradient of the $L_2$ norm is simply $\triangledown \|
\mathbf{x} \|_2 = \mathbf{x} / \| \mathbf{x} \|_2$, we have a bound on the
difference in the autocorrelation at time zero: 
\begin{align*}
|A_{\CoarseGraining}(L=0) - A(L=0)| \leq N \sqrt{|G|} \epsilon
  ~. 
\end{align*}
With increasing length we have:
\begin{align*}
A_{\CoarseGraining} (L) = \sum_i \pi_{\CoarseGraining} (i) c_i 
  \sum_{w \in \MeasAlphabet^L} \overline{f^{(w)}(c_i)} p^{(w)} (c_i)
\end{align*}
and:
\begin{align*}
A(L) & = \int_{\MxSSet} d\MxSMeasure(\mxst) \mxst
  \sum_{w \in \MeasAlphabet^L} \overline{f^{(w)} (\mxst)} p^{(w)} (\mxst) \\
  & = \sum_i \pi_{\CoarseGraining} (i)
  \int_{\mxst \in \CoarseGraining_i} d\MxSMeasure(\mxst | i) \mxst 
  \sum_{w \in \MeasAlphabet^L}
  \overline{f^{(w)} (\mxst)} p^{(w)} (\mxst)
  ~.
\end{align*}
Let $\mxst = c_i + \mathbf{\delta}$ for some mixed state in cell $i$. Then
we can write:
\begin{align*}
|A_{\CoarseGraining}(L) - A(L)| \leq \ \sum_i \pi_{\CoarseGraining} (i) 
 \left[  c_i \sum_{w \in \MeasAlphabet^L} \overline{f^{(w)}(c_i)} p^{(w)} 
 (c_i) - (c_i + \mathbf{\delta}) 
 \sum_{w \in \MeasAlphabet^L} \overline{f^{(w)}(c_i + \mathbf{\delta})}
  p^{(w)} (c_i + \mathbf{\delta}) \right]
  ~. 
\end{align*}
Now, note that:
\begin{align*}
p^{(w)} (c_i + \mathbf{\delta})
 \approx p^{(w)} (c_i ) + \triangledown p^{(w)} (c_i) \cdot \mathbf{\delta}
\end{align*}
and:
\begin{align*}
f^{(w)}(c_i + \mathbf{\delta}) \approx f^{(w)}(c_i) + e^{\lambda^x} 
 \mathbf{\delta}
\end{align*}
where $\lambda^w$ is the leading Lyapunov exponent of the mapping function. 
Substituting this and eliminating terms of order $\delta^2$ gives us:
\begin{align*}
|A_{\CoarseGraining}(L) - A(L)| \leq \ \sum_i \pi_{\CoarseGraining} (i) 
 \left[  c_i \sum_{w \in \MeasAlphabet^L} \left( \overline{f^{(w)}(c_i)} \triangledown p^{(w)} 
 \cdot \mathbf{\delta} +  e^{\lambda^{w}} \overline{\mathbf{\delta}} p^{(w)} 
 (c_i) \right)
 +
 \mathbf{\delta}
 \sum_{w \in \MeasAlphabet^L} \overline{f^{(w)}(c_i)}
  p^{(w)} (c_i) \right]
  ~.
\end{align*}
These terms identify three sources of approximation error: (i) that due to
a difference in the probability distribution over symbols, (ii) that in the
mapping functions, and (iii) that from approximating the points at the center
of their cells. 

For the first, we note that total variation in the probability
distribution over symbols is bounded by the distance between the mixed
states at which the distributions are computed. So, for any two mixed states
in the same cell, $\| \Pr(X = x| \mxst ) - \Pr(X = x| \mxstalt) \|_{TV} \leq
\sqrt{G}\epsilon$. Then, the first term is the error due to the difference
in the expectation value of the next state, given that we have calculated
the probability distribution at $c_i + \delta$, rather than $c_i$. Using
H{\"o}lder's inequality, for two distributions over $P(X)$ and $Q(X)$, we may
say:
\begin{align*}
E[f]_Q  - E[f]_P &= \sum_x f(x) (P(x) - Q(x)) \\ 
\sum_x f(x) (P(x) - Q(x)) & \leq \sum_x f(x) | P(x) - Q(x) | \\
\sum_x f(x) | P(x) - Q(x) | & \leq \| f \|_p \| P - Q \|_q
 ~,
\end{align*}
where $1/p + 1/q = 1$. Setting $q=1$:
\begin{align*}
E[f]_Q  - E[f]_P \leq \| f \|_{\infty} \| P - Q \|_{TV}
  ~.
\end{align*}
So, after taking the product with the cell centers $c_i$, we have that the
first error is bounded by $N \sqrt{G}\epsilon$ at all lengths. 

For the second, we note that since the maps are contractions, $\lambda < 0$,
and the distance between $f^x(\mxst)$ and $f^x(\mxstalt)$, where $\mxst$ and
$\mxstalt$ are in the same cell $i$, is bounded by $\sqrt{G}\epsilon$. As the
length of a word $w$ grows, $\lambda^w \to -\infty$ and the distance $f^w(
\mxst) - f^w(\mxstalt) \to 0$. At large $L$, this term vanishes, at a rate
equal to the average maximal Lyapunov exponent of the IFS. 

The final error is that in the autocorrelation in the cell approximation which
is, likewise, bounded by the cell size---this is the same error from $A(0)$,
viz. $N \sqrt{G}\epsilon$. 

And so, in combination with the bound on the entropy, we may say, loosely
speaking, that the error in the autocorrelation vanishes as $\epsilon \to 0$.
Therefore, to find $\tau$ and estimate the error in \cref{eq:MarkovAverage} as
a function of sample size, we take finer coarse-grained approximations until
convergence in the autocorrelation curve is observed, and then calculate $\tau$
directly. 

\end{document}